# Metabolic response to point mutations reveals principles of modulation of *in vivo* enzyme activity and phenotype

Sanchari Bhattacharyya[a], Shimon Bershtein[b], Bharat V. Adkar[a], Jaie Woodard[a] and Eugene I. Shakhnovich[a*]

[a]Department of Chemistry and Chemical Biology, Harvard University, 12 Oxford St, Cambridge, MA 02138
[b]Department of Life Sciences, Ben-Gurion University of the Negev, POB 653, Beer-Sheva 8410501, Israel

*Correspondence should be addressed to E.I.S. (shakhnovich@chemistry.harvard.edu)

## Abstract

The relationship between sequence variation and phenotype is poorly understood. Here we use metabolomic analysis to elucidate the molecular mechanism underlying the filamentous phenotype of *E. coli* strains that carry destabilizing mutations in the Dihydrofolate Reductase (DHFR). We find that partial loss of DHFR activity causes SOS response indicative of DNA damage and cell filamentation. This phenotype is triggered by an imbalance in deoxy nucleotide levels, most prominently a disproportionate drop in the intracellular dTTP. We show that a highly cooperative (Hill coefficient 2.5) *in vivo* activity of Thymidylate Kinase (Tmk), a downstream enzyme that phosphorylates dTMP to dTDP, is the cause of suboptimal dTTP levels. dTMP supplementation in the media rescues filamentation and restores *in vivo* Tmk kinetics to almost perfect Michaelis-Menten, like its kinetics *in vitro*. Overall, this study highlights the important role of cellular environment in sculpting enzymatic kinetics with system level implications for bacterial phenotype.

# Introduction

Understanding genotype-phenotype relationship is a central problem in modern biology. Mutations affect various layers of cellular organization, the mechanistic details of which remain far from being understood. Mutational effects propagate up the ladder of cellular organization from physico-chemical properties of biomolecules, up to cellular properties by altering the way proteins and nucleic acids function and interact with other cellular components. At the next level of hierarchy, mutations affect systems level properties, like the epigenome, transcriptome, proteome, metabolome or the microbiome. Collectively all layers in this multi-scale genotype-phenotype relationship dictates the fitness/phenotypic outcome of the mutations at the organism level. It has been shown , using the concept of a biophysical fitness landscape, that it is possible to predict fitness effects of mutations from a knowledge of molecular and cellular properties of biomolecules [1-4] as well as using systems level properties like proteomics and transcriptomics [5,6]. The metabolome which is represented by the metabolite profile of the cell, is a more recent advancement in the -omics technology [7]. Metabolites represent end products of biochemical pathways; hence they are downstream to other -omics data, and therefore closest to the phenotype. Hence metabolomics is widely recognized now as an important stepping-stone to relate genotype to phenotype [8-16]. In the recent past, high-throughput studies have been dedicated to understanding how genetic variations lead to changes in metabolic profile of the cell [17,18]. Though vast knowledge is available in terms of how mutations perturb metabolite levels either in the local vicinity or distant in the network, a mechanistic knowledge of how such changes modulate phenotypic outcomes is lacking.

In this work, we use targeted metabolomics to understand the mechanistic basis of how destabilizing mutations in the essential core metabolic enzyme of *E. coli* Dihydrofolate Reductase cause pronounced (>10 times of the normal cells) filamentation of bacteria. DHFR catalyzes conversion of dihydrofolate to tetrahydrofolate, which is an essential one-carbon donor in purine, pyrimidine and amino acid biosynthesis pathways. Metabolomics analyses reveal that filamentous mutant DHFR strains incurred a sharp drop in thymidine mono-, di- and tri-phosphate (dTMP, dTDP and dTTP), the latter being a deoxyribonucleotide that is essential for DNA synthesis and cell division. This results in DNA damage, upregulation of SOS response, and filamentation. We found that even though mutant strains have low dTMP levels, the disproportionately low levels of dTDP (and hence dTTP) arise primarily due to the strongly cooperative *in vivo* activity of the downstream essential pyrimidine biosynthesis enzyme Thymidylate Kinase (Tmk), which phosphorylates dTMP to dTDP. This is in stark contrast to its Michaelis-Menten (MM) activity profile observed *in vitro*. Surprisingly, supplementation of external dTMP in the medium which rescues *in vivo* dTDP levels and filamentation, switches the *in vivo* Tmk activity curves to the conventional '*in vitro* like' MM kinetics. The cooperative enzyme activity is best explained by the fractal nature of Tmk activity *in vivo* due to diffusion-limitation of substrate dTMP, possibly due to substrate channeling and metabolon formation. Overall, this study highlights the pleiotropic nature of mutations and way in which the complex cellular environment and metabolic network modulates *in vivo* enzyme activity and organismal fitness.

## Results

### *Several chromosomal mutations in folA gene give rise to slow growth and filamentation of E. coli*

Earlier we had designed a group of highly destabilizing chromosomal DHFR mutations in *E. coli* MG1655 (W133V, V75H+I155A, I91L+W133V, and V75H+I91L+I155A) that cause very slow growth at 37°C and 42°C ([19,20] and Figure 1A, see Methods for details about the strains). To understand the effects of these mutations on bacterial morphology, we grew mutant DHFR strains under two different growth conditions: M9 minimal media without and with supplementation with casamino acids (mixtures of all amino acids, except tryptophan, see *Methods*). In minimal media, median cell lengths of some mutants (W133V and V75H+W133V at 42°C) were smaller compared to WT, while I91L+W133V (at 40°C) was marginally longer than WT (Figure 1B). However, when M9 minimal medium was supplemented with amino acids, we found that cells carrying these mutations were pronouncedly filamentous. Figure 1C, D shows live cell DIC images of wild-type (WT) and I91L+W133V mutant DHFR strains at 30°C, 37°C, and 42°C (40°C for I91L+W133V strain). (See Supplementary Figure S1 for images of other low fitness mutant strains). In parallel to the detrimental effect of temperature on fitness, we noted that the morphologies were also temperature-sensitive. I91L+W133V and V75H+I91L+I155A strains exhibited a 1.5-1.75 fold increase (comparatively to WT) in the average cell length at 37°C (Figure 1E), while W133V and V75H+I155A were not elongated at 37°C. The latter, however, showed an increase up to 2.0-2.3 fold over WT cell lengths at 42°C (Figure 1F). Strains I91L+W133V and V75H+I91L+I155A showed 1.8-2.0 fold increase in the average cell length at 40°C, with some cells reaching up to 20μm in length

(about 10 fold increase) (Figure S1). Besides temperature of growth, since filamentation was also strongly dependent on availability of amino acids in the growth medium, it seemed likely that it was the result of a metabolic response due to partial loss of DHFR function.

*Filamentation is due to drop in DHFR activity*

DHFR is a central metabolic enzyme that is involved in conversion of dihydrofolate to tetrahydrofolate, and the latter is an important 1-carbon donor in the biosynthesis of purines, pyrimidines, and certain amino acids like glycine and methionine. Earlier we had reported that these mutant DHFR strains had very low abundance of the mutant proteins in the cell ([6,19] and Figure S2), an effect that could be rescued by deletion of Lon protease or by over-expressing chaperones like GroEL-ES [19]. We, therefore, reasoned that filamentation could be a result of drop in DHFR activity in these cells. To confirm this, we supplemented the *E. coli* strains carrying chromosomal DHFR mutations with WT DHFR expressed from a plasmid and found that both filamentation (Figure 2A) and growth defects (Figure 2B) were fully rescued. On the WT background, expression of extra DHFR resulted in some elongation, presumably due to toxicity of DHFR over-expression [15]. We also found that plasmid expression of mutant proteins in WT cells did not result in any filamentation (Figure 2C) or growth defects (Figure 2D). This shows that filamentation is not due to toxicity of the mutant DHFR proteins. We also found that treatment of WT cells with Trimethoprim, an antibiotic that targets bacterial DHFR, also caused filamentation at concentrations near the MIC (1µg/ml). At higher concentrations of the drug, there is growth arrest, hence no filamentation, leading to a non-monotonic dependence of cell length on TMP concentration (Figure S4A, B at 37°C and 42°C respectively).

*Filamentous strains exhibit imbalance between dTTP and other deoxyribonucleotides*

Here we aimed to determine metabolic changes in mutant strains associated with filamentous phenotype. To that end we carried out metabolomics analysis of mutant strains under conditions of filamentation (in amino acid supplemented M9 medium at 42°C for WT and W133V, at 40°C for I91L+W133V, at lower concentration of TMP, close to MIC which exhibited pronounced elongation phenotype) as well as under non-filamentation conditions (in minimal medium at 42°C for WT and W133V and at 40°C for I91L+W133V and in the absence of amino acids for all strains).

We observed that in the absence of amino acids, *when the cells are not filamented* mutant strains as well as WT cells treated with 1µg/ml TMP (close to MIC) exhibited very low levels of both purines and pyrimidines (Figure 3A, Figure S5A, B; Table S1). For example, in strain I91L+W133V, IMP, AMP and dTMP levels were respectively 17%, 30% and 5% of WT levels, while dTTP levels were below the detection limit. Methionine and glycine biosynthesis require tetrahydrofolate derivatives, hence, expectedly, methionine levels were only 1-3% in mutant strains (Figure 3B, Figure S5A,B). Overall, we conclude that large drop in methionine and purine (IMP) levels presumably stalls protein/RNA synthesis. Since increase in cell mass is essential for filamentation, cells under this condition are not filamented.

In the presence of 1% casamino acids, several but not all amino acids showed a marked increase in abundance (Figure 3D, Figure S5 and Table S1). Methionine levels rose to 40% of WT levels for I91L+W133V mutant, while aspartate/asparagine, glutamine, histidine and tryptophan levels also showed a significant increase. Particularly interesting was the fact that purine levels were substantially rescued upon addition of amino acids (Figure 3C, Figure S5A,B). IMP showed the maximal effect, increasing 10-15 fold over its levels in the absence of amino acids. ATP, ADP, AMP and GMP also showed similar trends. Since the product of DHFR is eventually used in the synthesis of methionine, IMP and dTMP (Figure S6A), we hypothesize that addition of methionine in the medium allows higher amounts of 5,10-methylene-THF to be channeled towards synthesis of purines and pyrimidines. Moreover, both de novo purine and pyrimidine biosynthesis pathways require aspartate and glutamine (Figure S6B, C), which were otherwise low in minimal medium. Overall, the metabolomics data suggest that in the presence of added amino acids, protein, and RNA synthesis are no longer stalled, and therefore growth, which is pre-requisite for filamentation, can happen.

Though dTMP levels increased to about ~10% of WT levels in I91L+W133V and ~50% in W133V, surprisingly, dTTP levels (thymidine derivative that is incorporated in the DNA) were only about 1% of WT levels in I91L+W133V and 18% for W133V (Figure 3C, Figure S5A, B). In contrast, dATP and dCTP levels were very high (Figure 3C, Figure S5A, B). We hypothesize that as cellular growth continues, misbalance in the concentrations of deoxy nucleotides may lead to erroneous DNA replication, induction of SOS response and blocked cell division. Indeed, in our previous study, proteomics and transcriptomics analyses showed that several SOS response genes were upregulated in I91L+W133V and V75H+I91L+I155A strains at 37°C [6].

*Filamentation and SOS response: Deletion of recA rescues filamentation*

We quantified the expression of several SOS genes *recA*, *recN* and *sulA* under different supplementation conditions for mutant DHFR strains as well as TMP-treated WT cells. Mutants grown at low temperature and those grown in the absence of amino acids were not elongated, and not surprisingly, they did not elicit any SOS response (Figure 4A-C). In comparison, for all conditions that are associated with filamentation (I91L+W133V mutant at 37°C, W133V and I91L+W133V at 42°C) induced strong SOS response (Figure 4A-C). The levels of induction of these genes were greatly reduced in the presence of dTMP, consistent with lack of filamentation under this condition (Figure 4A-C). Similar trends were also observed for WT treated with TMP.

Overexpression of RecA is the key trigger of the SOS response to DNA damage. RecA cleaves the dimeric LexA repressor to turn on the genes that are under the SOS box (e.g., sulA, uvr proteins, *etc.*). SulA inhibits the cell division protein FtsZ, eventually causing filamentation. To understand the role of *recA* and *sulA* in our study, we treated *ΔrecA* and *ΔsulA* strains with near-MIC levels of the antibiotic TMP. As expected, *ΔrecA* strain did not show filamentation upon TMP treatment (Figure 4D), clearly highlighting its definitive role in filamentation. However, *ΔsulA* strain continued to filament (Figure 4D), suggesting possible role of sulA-independent pathways. This data, however, does not negate out the role of sulA, since sulA is highly upregulated in mutant strains and is a well-known inhibitor of the cell division protein FtsZ. A poorly characterized segment of the *E. coli* genome called the e14 prophage that harbors the *sfiC* gene has been implicated to play a role in SOS dependent but sulA independent filamentation [21]. We found that a mutant *E. coli* has been knocked out for the e14 region does not filament upon TMP treatment (Figure 4D), indicating that *sfiC* might be one of the players involved in elongation. Overall, these results clearly establish the role of the SOS pathway towards filamentation in mutant DHFR strains, which in turn is due to imbalance between dTTP and other deoxynucleotides in the cell.

*Filamentation in mutant DHFR strains is not due to TLD and is reversible*

To find out if mutant strains represented a case of thymineless death (TLD), a condition that causes extensive filamentation due to extreme thymine deprivation [22], we assessed the viability of mutant strains at 30°C (permissive temperature) on solid media after several hours of growth at 42°C (restrictive temperature) in the presence of amino acids (filamentation condition). Sangurdekar *et al* [23] reported exponential loss in viability in TLD after one hour of growth under thymineless conditions. The idea was that if cells underwent TLD under conditions of filamentation, they would no longer be able to resume growth/form colonies at permissive temperature. To that end, we induced filamentous phenotype by incubating W133V mutant cells for 4 hours at 42°C, and then monitored cell recovery at room temperature (see *Methods*). Figure 5A shows a representative example of morphology of W133V filament that begins to undergo slow division initiated at the poles at permissive temperature. Most progeny cells of normal size appeared after 5-6 hours of growth at low temperature, indicating no loss in viability.

Earlier studies have shown that inhibition of DHFR activity by Trimethoprim (TMP) under conditions of both amino acids and purine supplementation leads to TLD [24]. We found that WT cells that were subjected to *very high* TMP concentrations showed loss of colony forming units on media supplemented with *both* amino acids and a purine source (GMP), indicating death (Figure 5C), similar to [24], while supplementation with only amino acids was bacteriostatic (Figure 5B). Since mutant DHFR strains incur only *partial* loss of DHFR activity, they resemble lower concentrations of TMP treatment, and therefore did not show any substantial loss in colony forming units when grown in the presence of amino acids and GMP (Figure 5D). Collectively, these experiments demonstrated that despite extensive DNA damage and filamentation, mutant cells did not represent a case of TLD.

### *Supplementation of dTMP in the medium restores dTDP/dTTP levels and rescues filamentation*

Since mutant strains had very low dTMP/dTTP levels, we grew WT and mutant strains in minimal medium that was supplemented with both amino acids and 1mM dTMP and carried out measurement of cell length as well as metabolomics analysis. Addition of dTMP largely rescued filamentation of mutant strains (Figure 6A). Metabolomics analyses showed that under this condition, I91L+W133V mutant had much higher levels of both dTDP and dTTP relative to WT (Figure 6B). Concentration of the other deoxyribonucleotides dATP/dCTP levels however, remained high despite addition of dTMP (Figure 6B). Therefore, we conclude that higher amounts of dTTP presumably reduce the imbalance in relative concentrations of deoxy-nucleotides, thus relieving DNA damage and filamentation (see above).

Moreover, supplementation of dTMP and amino acids also allowed mutants W133V, I91L+W133V and V75H+I91L+I155A to form higher counts of colony forming units (cfu) than with only amino acids at their respective filamentation temperatures (Figure 6C shows images for V75H+I91L+I155A at 40ºC). However, we note that cfu count for mutants in the presence of dTMP were still orders of magnitude lower than those of WT, indicating that thymine only rescues cell length, and not growth defects. This was also supported by the absence of growth rate rescue with dTMP (Figure 6D).

*Low dTDP/dTMP ratio: possibility of inhibition of Thymidylate Kinase?*

Interestingly, we found that while dTMP levels are low in mutant DHFR strains (10% and 50% of WT levels in I91L+W133V and W133V mutants respectively), dTDP levels were far lower (1% and 20%) (Figure 3C, Figure S5A). On the other hand the ratios of dTDP to dTTP were approximately equivalent in both WT and mutants. Supplementation of dTMP in the medium however, restores the relative abundances of dTMP to dTDP to dTTP to approximately WT level in I91L+W133V (Figure 6B). It raises the possibility that the pyrimidine biosynthesis pathway enzyme Thymidylate Kinase (Tmk), which phosphorylates dTMP to dTDP, might be inhibited in mutant DHFR strains. Previous reports suggest that dUMP and dCTP can act as competitive inhibitors for Tmk [25], and interestingly, we found that both dUMP and dCTP levels were highly upregulated in mutant cells (Figure 3C), due to inefficient conversion of dUMP to dTMP when DHFR activity was reduced (Figure S6C). To find out if intracellular accumulation of dUMP and dCTP in mutant cells is sufficient to inhibit Tmk, we overexpressed and purified his-tagged Tmk from *E. coli* cells and tested the potential inhibitory effect of dUMP and dCTP *in vitro*. In comparison to its cognate substrate dTMP for which the $K_M$ is 13µM (Figure 7A), the $K_M$ for dUMP is 450µM (Figure S7A), which indicates that dUMP is a much weaker substrate as compared to dTMP. The apparent $K_I$ of dUMP for Tmk was 3.9mM (Figure S7B). Given that dUMP concentration inside WT *E. coli* cells is 0.01mM [26], its concentration inside mutant DHFR cells would be in the range of 0.5mM (50-fold higher levels), which is not enough to cause substantial inhibition of Tmk *in vivo*. We next carried out an activity assay of Tmk in the presence of varying amounts of dCTP. Depending on magnesium concentration, the $K_I$ of dCTP ranged between 2-4 mM (Figure S7C). Considering intracellular dCTP concentration in WT cells to be 35µM [7], those in mutant cells are about 0.7mM (20-fold overexpression). Hence, like dUMP, intracellular dCTP levels are too low to show any substantial inhibition of Tmk activity *in vivo*. To conclude, though dUMP and dCTP have the potential to inhibit Tmk activity, their concentrations inside mutant cells are not high enough to achieve that.

*Cooperative of enzymatic activity of Tmk in vivo explains low dTDP levels*

As discussed previously, our metabolomics data shows that in mutant cells, dTDP levels fall far more precipitously than dTMP concentrations. If Tmk follows the same Michaelis-Menten (MM) like dependence on intracellular dTMP concentrations as seen *in vitro*, we find that dTDP levels would never drop as low as the experimentally observed value for mutants (Figure 7B, black line for I91L+W133V) for any assumed intracellular dTMP concentration for WT. To resolve this inconsistency, we attempted to elucidate the *in vivo* activity curve of Tmk enzyme from dTDP and dTMP levels measured from the metabolomics data for WT and mutant cells. To get a broad range of data, we measured metabolite levels for WT and several mutants at different time points during growth both in the absence and presence of different concentrations of external dTMP in the medium. Surprisingly, we found that the data points from our metabolomics experiments traced two different curves depending on whether there was external dTMP in the medium. Data points derived in the *absence* of external dTMP had a long lag, followed by a more cooperative increase (red points in Figure 7C), while the data points corresponding to added dTMP, appeared to follow the traditional MM curve (black data points in Figure 7C) and was similar to the enzyme activity of Tmk observed *in vitro*. We fitted both datasets to a Hill equation. For the conditions without added dTMP (red data points of Figure 7C) Hill-coefficient of 2.5 was obtained suggesting strong positive cooperativity. On the other hand, when dTMP was added, the dTDP vs dTMP curve (black datapoints in Figure 7C) was fitted with a Hill coefficient of 1.2. However, the fit was not significantly different from the traditional MM model (p-value = 0.36). Based on the Hill-like curve in Figure 7C, we can say that for intracellular dTMP concentrations below 10μM, a 10% drop in dTMP levels (as seen for I91L+W133V) would cause dTDP levels to drop to 1% or less (Figure 7B, red line), largely because of the long lag. Therefore, a Hill-like dependence of Tmk activity on dTMP concentrations can explain the disproportionately low dTDP levels in mutant strains.

However, we note that metabolomics does not directly report on the kinetics of an enzyme *in vivo*. Rather, it reports on the steady state levels of metabolites present inside the cell at any given time. Moreover, unlike *in vitro* activity measurement conditions where an enzyme functions in isolation, *in vivo* Tmk is a part of the pyrimidine biosynthesis pathway that involves several sequentially acting enzymes as follows:

$$E_1 + S_1 \underset{k_{-1}}{\overset{k_1}{\rightleftharpoons}} E_1S_1 \underset{k_{-2}}{\overset{k_2}{\rightleftharpoons}} E_1 + S_2$$
$$E_2 + S_2 \underset{k_{-3}}{\overset{k_3}{\rightleftharpoons}} E_2S_2 \underset{k_{-4}}{\overset{k_4}{\rightleftharpoons}} E_2 + S_3 \rightleftharpoons \ldots\ldots$$
(1)

Where all metabolites $(S_1, S_2, S_3,..)$ are at steady state. Assuming that each enzyme in this pathway follows Michaelis-Menten kinetics and that the *product* of each enzyme has very low affinity to bind the enzyme back, we arrive at the following equation (see detailed derivation in the Supplementary Text):

$$[S_2] = \frac{k_2 K_{M2}[E_1]_0[S_1]}{k_4 K_{M1}[E_2]_0 + (k_4[E_2]_0 - k_2[E_1]_0)[S_1]} = \frac{A[S_1]}{B + C[S_1]} \tag{2}$$

Where $A = k_2 K_{M2}[E_1]_0$, $B = k_4 K_{M1}[E_2]_0$, $C = (k_4[E_2]_0 - k_2[E_1]_0)$ and $K_{M1}$ and $K_{M2}$ are the Michaelis constants of enzymes $E1$ and $E2$ for $S1$ and $S2$, $[E_i]_0$ is a total concentrations (i.e. free+bound) of an i-th enzyme. In other words, the steady state concentration of any product in the pathway follows a hyperbolic or MM like dependence on its substrate concentration, similar to the black curve (Figure 7C) obtained in the presence of external dTMP.

Next, we assume that Tmk (along with other enzymes in this pathway) follows a Hill-like kinetics (due to reasons we elaborate in the Discussion) in the following form:

Initial rate $v_0 = \frac{v_{max}[S]^m}{K_M + [S]^m}$, where '$m$' is the Hill coefficient,

then for the pathway of enzymes at steady state (as in Eq (1)), $S2$ has the following dependence on $S1$:

$$[S_2]^n = \frac{k_2 K_{M2}[E_1]_0[S_1]^m}{k_4 K_{M1}[E_2]_0 + (k_4[E_2]_0 - k_2[E_1]_0)[S_1]^m} = \frac{A[S_1]^m}{B + C[S_1]^m},$$ where '$m$' and '$n$' are the Hill coefficients of consecutive enzymes $E1$ and $E2$.

Hence, $[S_2] = \left[\frac{A[S_1]^m}{B + C[S_1]^m}\right]^{1/n}$ (3)

As shown in Supplementary Text, the above equation gives rise to a Hill like dependence of $S2$ on $S1$ with positive cooperativity, similar to the red curve (Figure 7C) obtained in the *absence* of external dTMP, with the assumption that $m > n$. The above analysis suggests that the dependence of the steady state concentrations of metabolites along the linear pathway on each other is reflective of the kinetics of the concerned enzyme *in vivo*, hence it is reasonable to infer that *in vivo* kinetics of Tmk is Hill-like under native conditions and MM like in the presence of added dTMP.

*Supplementation of thymidine retains cooperative behavior of Tmk in vivo*

dTMP, the substrate of Tmk, comes from two different sources inside the cell: the *de novo* pyrimidine biosynthesis pathway through conversion of dUMP to dTMP by thymidylate synthase, and the pyrimidine salvage pathway through conversion of thymine to thymidine to dTMP. Mutant DHFR strains which are unable to efficiently convert dUMP to dTMP through the *de novo* pathway due to reduced folate activity, rely substantially on the salvage pathway for their dTMP supply, as has been shown for catalytically inactive mutants of DHFR [14]. Hence, we next asked the question: what happens to the Tmk activity curve *in vivo* if dTMP is produced (largely) through the salvage pathway instead of being directly supplied from an external source? To that end, we supplemented the growth medium with intermediates from the salvage pathway, namely thymine and thymidine. While supplementation of thymidine increased intracellular dTMP levels for WT as well as mutants (Figure S8A), the dTDP vs dTMP levels followed the Hill like curve (Figure 7D, inset figure shows data for the I91L+W133V mutant in the absence of supplementation and in the presence of added thymidine and dTMP). This shows that direct supplementation of the substrate (dTMP) of Tmk results in very different enzyme kinetics compared to when a precursor of dTMP is supplied externally.

Quite surprisingly and contrary to WT, *mutant* strains did not use external thymine base towards increasing intracellular dTMP (Figure S8A), though it was uptaken by the cells (Figure S8B). We ruled out inhibition of DeoA enzyme (which interconverts thymine and thymidine) in mutants, as thymidine supplementation increases thymine levels significantly (Figure S8C). However, I91L+W133V mutant had considerably lower level of deoxy-D-ribose-1-phosphate (dR-1P) (Figure S8D) which is used as a substrate by the enzyme DeoA to synthesize thymidine from thymine. This strain also accumulated large excess of deoxy-D-ribose-5-phosphate (dR-5P) (Figure S8D), indicating that isomerization of the sugar dR-1P to dR-5P through DeoB enzyme might be one of the reasons for the lack of thymine utilization. This scenario is supported by the recent finding that cells that evolved to grow on small amounts of thymine supplement on the background of inactive DHFR mutationally deactivated DeoB thus blocking the channeling of dR-5P towards glycolysis and providing sufficient amount of dR-1P towards thymidine synthesis in the salvage pathway [14].

## Discussion

Metabolic networks of cells are inherently intertwined, with substrates and products of one pathway being utilized by another pathway. As a result, perturbations produced in one pathway can easily percolate into others, usually magnifying effects. The folate pathway or the 1-carbon metabolism pathway is a classic example of this, as reduced folates act as 1-carbon donors during biosynthesis of purines, pyrimidines and amino acids. Kwon et al [27] showed that for inhibition of DHFR activity using trimethoprim, accumulation of substrate dihydrofolate (DHF), in turn, results in inhibition of another downstream enzyme critical to folate metabolism: folylpoly-gamma-glutamate synthetase (FP-gamma-GS), in a domino like effect (falling DHFR activity triggers a fall in the other enzyme's activity too). In this work, we show that in *E. coli* strains that harbor destabilizing mutations in *folA* gene, reduced DHFR activity strongly affects, among other factors, the pyrimidine biosynthesis pathway by reducing production of dTMP from dUMP via thymidylate synthase (ThyA) that uses a derivative of THF as one carbon source. Much like a domino effect, such drop in dTMP levels due to mutations in DHFR results in a precipitous drop in dTDP/dTTP, mainly due to the strong cooperative *in vivo* activity of another downstream essential enzyme Thymidylate Kinase (Tmk) in the pyrimidine biosynthesis pathway. Drop in dTTP level eventually leads to an imbalance in the levels of deoxynucleotides, causing errors in DNA replication, SOS response and filamentation.

An important finding from the current study is that enzymes can exhibit a different kinetics *in vivo* depending on the source of the substrate. In case of Tmk, the enzyme showed a conventional Michaelis-Menten type *in vivo* activity when dTMP was externally supplied through the growth medium. However, when an equivalent concentration of dTMP was produced by the cell itself using its own cascade of enzymes in the pathway, it showed a dramatically different cooperative (Hill-like) activity. There are two important questions that arise out of this observation: first, why is the intrinsic *in vivo* activity of Tmk Hill-like? Second, what causes this shift from Hill-like to Michaelis-Menten (MM)? One of the most straightforward reasons for Hill-like enzyme activity is allosteric substrate binding. However, purified Tmk *in vitro* shows perfect MM kinetics, ruling out any intrinsic allostery of the enzyme. We also found that even in the presence of high concentrations of dUMP and dCTP (the known inhibitors of Tmk), the activity of Tmk conforms to MM kinetics ruling out these metabolites as allosteric regulators (Figure S7D). The other possible mechanism of Hill-like kinetics is 'limited diffusion' of one or more of the interacting components of a reaction [28,29]. Conventional

MM enzyme kinetics relies on the assumption of free diffusion, and hence laws of mass action are obeyed. However, in case of limited diffusion, conditions of spatial uniformity are no longer maintained; hence, law of mass action is not applicable. Theoretical work as well as simulations [28-33] have shown that such diffusion limited reactions often exhibit kinetics with Hill-like coefficients that are significantly higher than 1, and often fractional (so called fractal kinetics), similar to Hill coefficients obtained with our data (Figure 7C, Hill coefficient=2.5). It has also been postulated that biological reactions, especially those that happen in dimensionally restricted environments like 1D channels or 2D membranes exhibit fractal like kinetics [30,34]. In our case, it seems more reasonable that it is the substrate dTMP that has limited diffusion rather than the enzyme Tmk itself, since addition of external dTMP alleviates the Hill-like effect (in Supplementary Text, we show a derivation of Hill-like enzyme kinetics assuming that only the substrate is diffusion limited, using a power law formalism as developed by Savageau [29]). But why should dTMP be diffusion limited? Substantial work in the recent past has shown that metabolic enzymes of a pathway, including those involved in purine biosynthesis [35,36] as well as in 1-carbon metabolism [15] form a metabolon, a supramolecular complex comprised of transiently interacting enzymes, that allows efficient channeling of metabolites. Though channels help in easy exchange of metabolites between consecutive enzymes and prevent their unwanted degradation or toxicity in the cytosol, they have reduced dimensionality compared to the cytosol, thereby making motion less 'random' and hence limiting diffusion of the substrate/products. In our case, it is possible that enzymes of the pyrimidine biosynthesis pathway as well as the salvage pathway form a metabolon, that limits diffusion of dTMP. External dTMP on the other hand, is free to diffuse in the cytoplasm, and hence results in traditional MM kinetics to emerge with a Hill coefficient of 1.

Though we do not have direct evidence of Tmk being involved in a metabolon, our study does show some circumstantial evidence. Our previous work on DHFR showed that toxicity and filamentation upon over-expression might be a hallmark of metabolon proteins [15], through sequestration of neighboring/sequential proteins in the pathway. On a similar note, we found that overexpression of Tmk in WT *E. coli* cells led to filamentation (Figure S9), strongly suggesting that Tmk might be part of a metabolon. It is worth mentioning at this point that Tmk overexpression in mutant DHFR cells does not rescue filamentation (Figure S9). This is consistent with the metabolon hypothesis, since dTMP produced by the cell would still be confined to the metabolon, and hence overexpressed Tmk cannot overcome the problem of diffusion limitation of dTMP.

For decades, enzyme activity has been studied *in vitro* with purified enzyme in dilute solution with excess substrate. Though *in vitro* measured parameters have been largely successful to interpret cellular data [2,4], in other cases they have only provided limited information [37]. Substantial efforts in the recent past have therefore been directed towards replicating *in vivo* like conditions with purified enzymes [38-40]. These include macromolecular crowding, pH conditions and buffer capacity [41]. In this work, we show how the *in vivo* activity curves of an enzyme can be markedly different (in case of Tmk strongly Hill-like) from its perfect MM like kinetics *in vitro*. Based on available literature and some of our preliminary experiments, metabolon formation and subsequent diffusion limitation of substrate dTMP through the *de novo* and salvage pyrimidine biosynthesis pathway seems like the most probable mechanism. Future work will prove or disprove this hypothesis. However, regardless of the mechanism, this work provides convincing evidence that the cellular environment can modulate enzyme activity in a very fundamental way, which explains a key bacterial phenotype in our case.

In this study, we used metabolomics as the key tool to link molecular effects of mutations to phenotype and illustrate precise biochemical and biophysical mechanisms through which altered metabolite levels modulate bacterial phenotypes, in this case, filamentation. Detailed metabolomics analysis allowed us to pinpoint the pathway and specific enzyme responsible for the phenotype and, surprisingly, it turned out to be far downstream from the mutant locus (folA). Furthermore, the culprit, Tmk, does not use products of the folate pathway as a cofactor. Nevertheless, it appears that perturbation of the folate pathway caused by mutations in DHFR propagated downstream in a domino-like manner to create a bottleneck in a specific metabolite dTDP triggering cellular SOS response and pronounced phenotypic effects manifested in altered cell morphology. Altogether our results show how metabolomics can be used as a stepping-stone from biophysical analysis of variation of molecular properties of enzymes to phenotypic manifestation of mutations and close the gap in the multi-scale genotype-phenotype relationship.

## Acknowledgements

This work was supported by NIH grant GM068670 to E.I.S.

## Author contributions

Conceptualization, S.Bh., S.Be., and E.I.S.; Methodology, S.Bh., S.Be., B.V.A., J.W., and E.I.S.; Formal Analysis, S.Bh., S.Be., B.V.A., and E.I.S.; Investigation, S.Bh., S.Be, B.V.A., J.W.; Writing-

original draft, S.Bh., and E.I.S.; writing-review & editing, S.Bh., S.Be., B.V.A., and E.I.S.; Supervision, E.I.S.; Funding acquisition, E.I.S.

# Figures and Legends

Figure 1

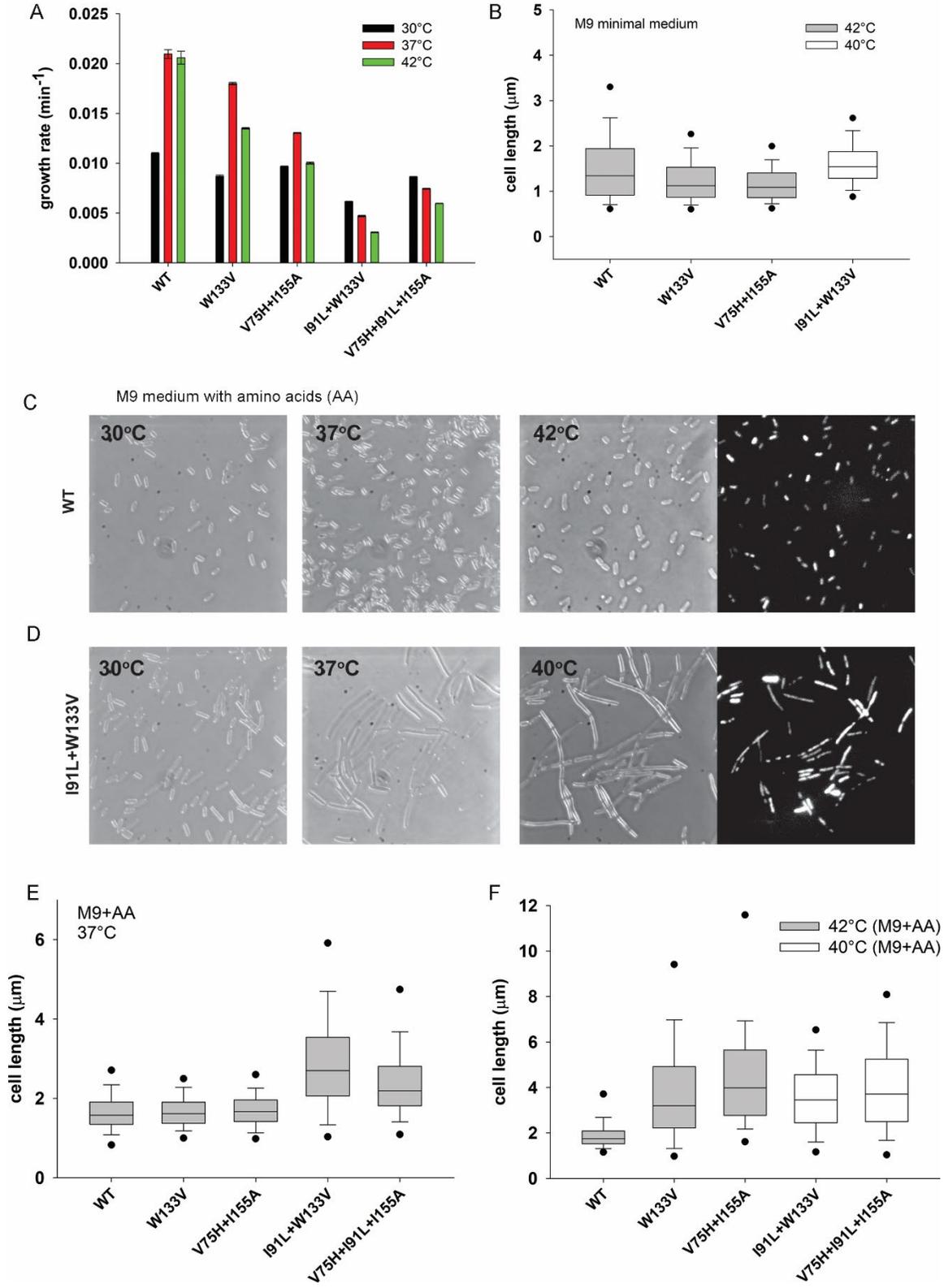

**Figure 1: Destabilizing mutations in DHFR induce filamentous phenotype.** (A) Growth rates of mutant DHFR strains at 30°C, 37°C and 42°C. While most mutants grow well at 30°C, they grow very poorly at high temperatures. Error bars represent SEM of three biological replicates. (B) Distribution of cell lengths of WT and mutants W133V and V75H+I155A at 42°C (gray box) and I91L+W133V at 40°C (represented as white box) after being grown in M9 minimal medium for 4 hours. Median cell length of W133V and V75H+I155A is significantly smaller than WT (Mann-Whitney test, p-value <0.001). Live cells DIC images and DAPI nucleoids staining of (C) WT DHFR and (D) I91L+W133V DHFR strains after being grown at 30°C, 37°C, or 42°C (I91L+W133V was grown at 40°C) in M9 medium supplemented with amino acids for 4 hours (see *Methods*). Cell lengths were measured from the obtained DIC images (see *Methods*) and their distribution at 37°C and 40°C/42°C is shown in (E) and (F) as box-plots *(see Methods)*. Images of other mutant DHFR strains W133V, V75H+I155A and V75H+I91L+I155A are presented in related Figure S1.

**Figure 2**

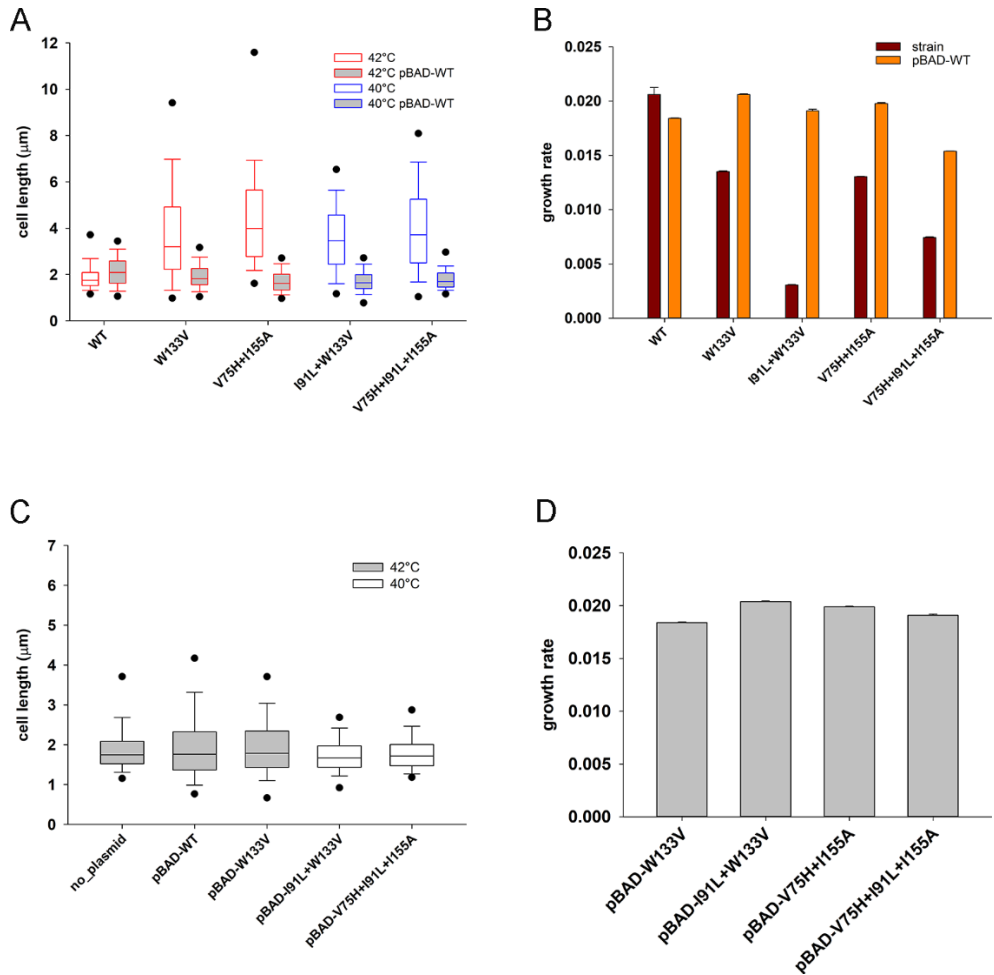

**Figure 2: Filamentation in mutant DHFR strains is due to loss of DHFR activity.** WT and mutant DHFR strains were transformed with pBAD plasmid that expressed WT DHFR under control of arabinose promoter. Transformed cells were grown at 42°C (for WT, W133V and V75H+I155A strains) or at 40°C (for I91L+W133V and V75H+I91L+I155A) in M9 medium supplemented with amino acids. Functional complementation of WT DHFR rescues both (A) filamentation and (B) growth defects of mutant strains. Expression of mutant proteins from pBAD plasmid on the WT background does not result in (C) filamentation or (D) growth defects. See related Figure S2.

Figure 3

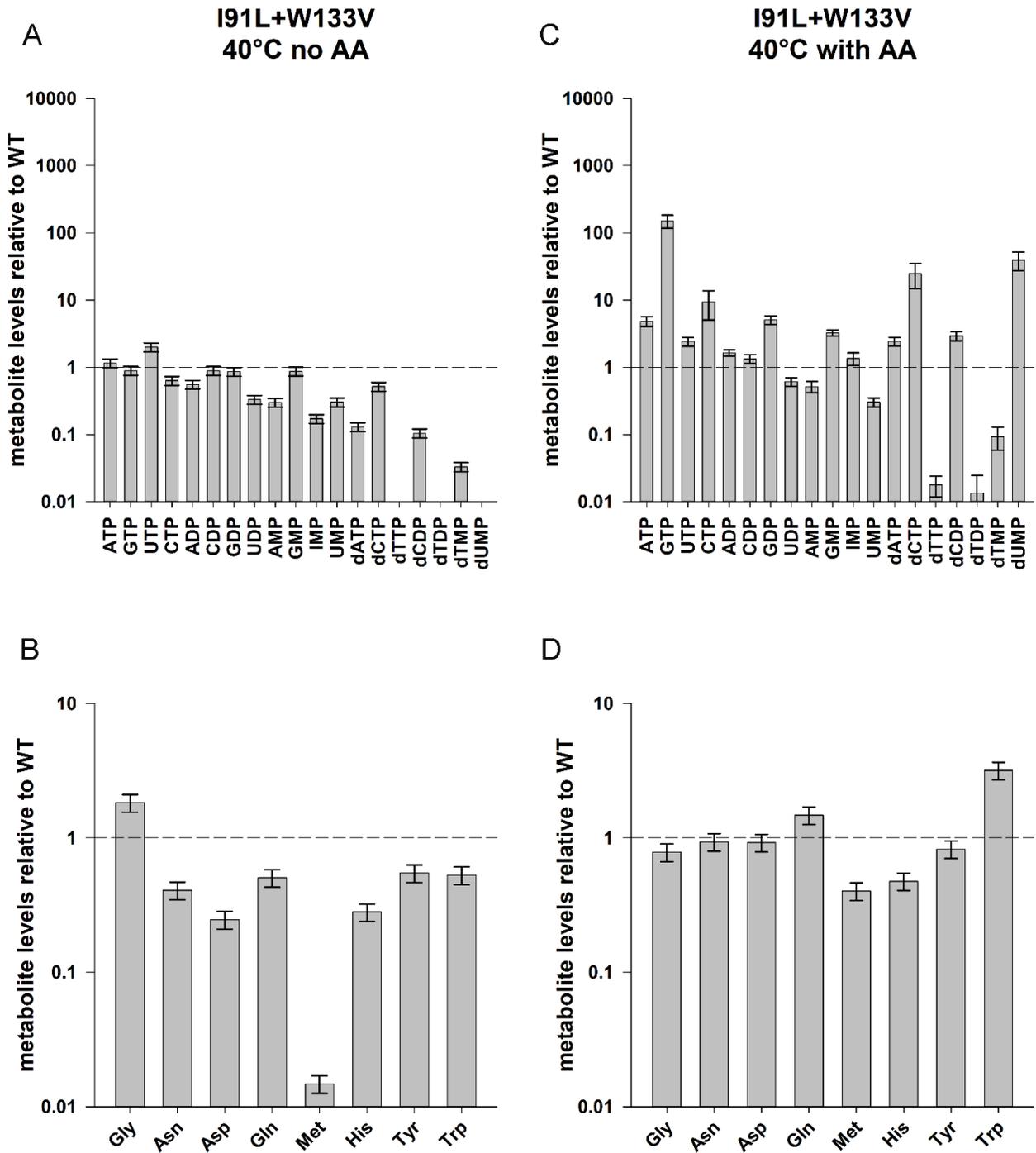

**Figure 3: Metabolomics of mutant DHFR strains in minimal media without or with added amino acids.** (A) and (B) shows abundance of selected nucleotides and amino acids for mutant I91L+W133V after 4 hours of growth at 40°C in M9 minimal medium (no filamentation), while (C) and (D) represents nucleotide and amino acid abundances after 4 hours of growth in amino

acid supplemented M9 medium at 40°C (condition of filamentation). Concentration of all metabolites were normalized to WT levels at 4 hours when grown under similar conditions. In minimal medium (B), Methionine levels are extremely low, which recover in panel (D). Levels of purines (IMP, AMP) as well as pyrimidines (dTMP) are rescued with amino acid supplementation, however dTDP and dTTP levels remain extremely low. Error bars represent SEM of at least three biological replicates (see *Methods*). See related Figures S5A-B and Table S1.

**Figure 4**

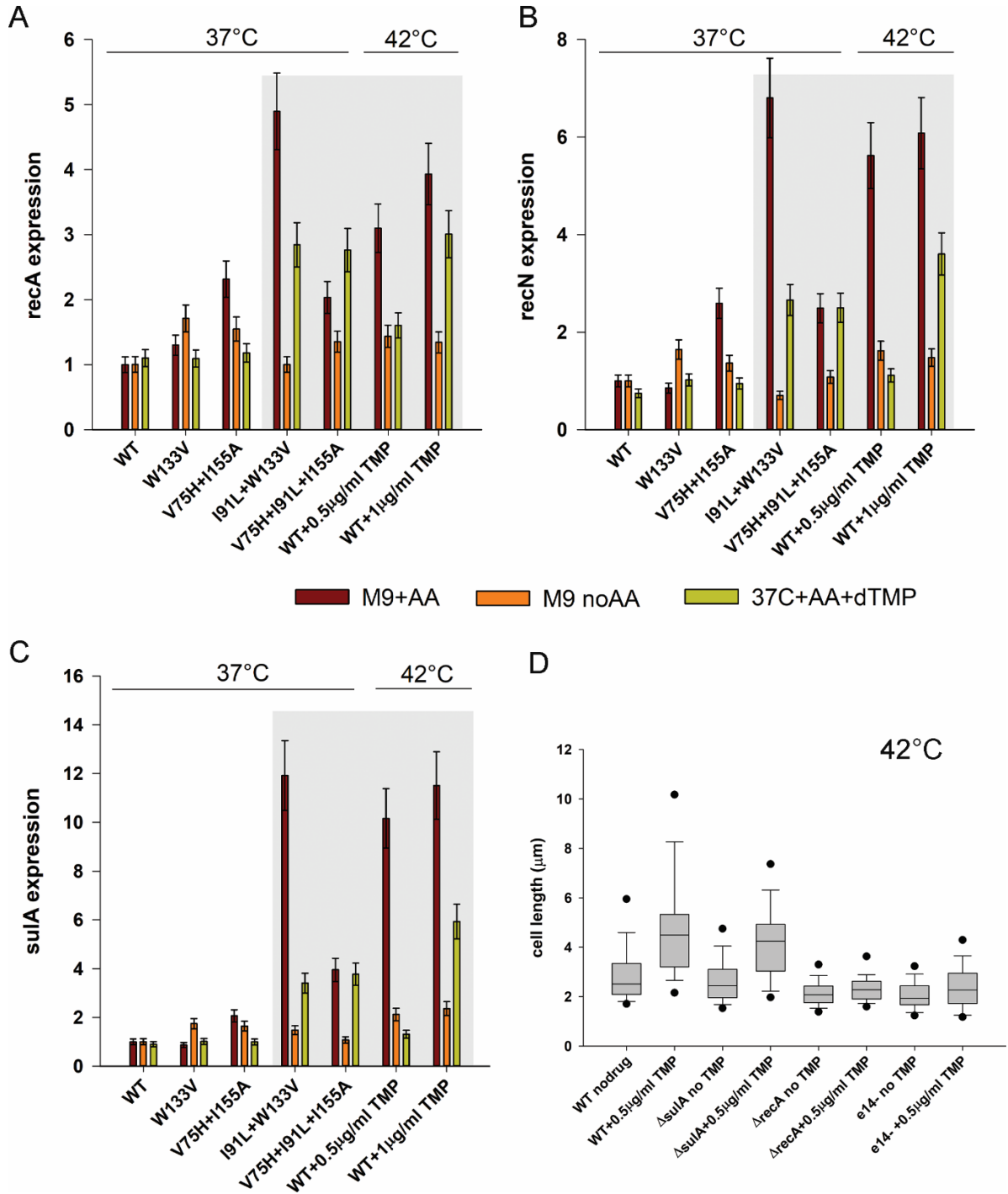

**Figure 4: Filamentation in mutant DHFR strains is associated with strong SOS response.** Expression of (A) *recA* (B) *recN* and (C) *sulA* genes measured by quantitative PCR when WT and mutant strains are grown in M9 medium with or without supplementation of amino acids or dTMP.

WT and mutant strains were grown for 4 hours of growth in the indicated medium at 37°C, while WT treated with different concentrations of TMP were grown for 4 hours at 42°C. Brown bars (M9+AA) in the gray shaded area correspond to filamentation conditions and these are associated with pronounced upregulation of all three SOS genes. On the other hand, conditions with loss of filamentation (with dTMP or no supplementation) show much less expression. Error bars represent SD of 2-3 biological replicates (see *Methods*). (D) Treatment of WT *E. coli* cells with sub-MIC concentration of TMP (0.5µg/ml) leads to filamentation at 42°C when grown in amino acid supplemented medium. However, a *recA* knock-out strain under similar condition shows no elongation, indicating the role of SOS pathway in filamentation. A *sulA* knock-out continues to elongate, indicating the role of *sulA*-independent pathways. An *E. coli* strain deleted for the e14 prophage region however showed no filamentation upon TMP treatment, indicating that sfiC gene in the e14 region might be one such sulA independent player.

**Figure 5**

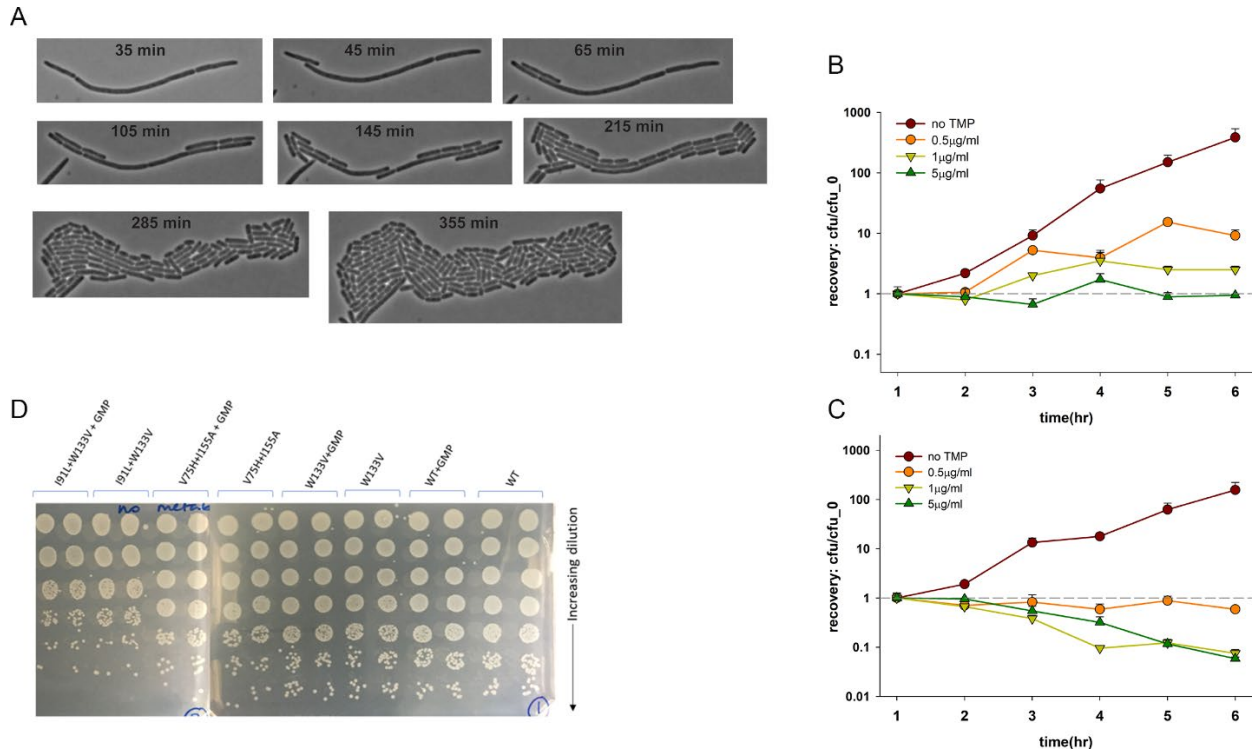

**Figure 5: Filamentation in mutant DHFR strains is completely reversible.** (A) Mutant W133V was grown in amino acid supplemented M9 medium (M9+AA) for 4 hours at 42°C, and subsequently placed on M9 agar pads and their growth was monitored at room temperature. Shown are phase contrast images taken from different time points throughout the time-lapse experiment. Unlike cells experiencing TLD, an irreversible phenomenon, W133V DHFR cells recover and resume growth at low temperature. (B and C) WT cells were treated with different concentrations of TMP at 42°C for varying amounts of time in amino acid supplemented M9 medium (panel B) or in M9 media supplemented with both amino acids and GMP (panel C), following which they were spotted on M9+AA plates and allowed to grow at 30°C. Colonies were counted next day. In the presence of only amino acids, there was no loss in viability for any concentration of TMP (panel B), despite extensive filamentation (Figure S4B). In contrast, in the presence of amino acids and GMP, the cells showed sharp loss in viability when grown at high TMP concentrations. In both panels, error bars represent SD of three biological replicates. (D) WT and mutants were grown as in (A) for 6 hours at 42°C in M9+AA medium without or with GMP, and subsequently diluted serially and spotted on M9+AA agar plates and allowed to grow at 30°C till visible colonies were formed. No loss in viability was observed for WT or mutants.

**Figure 6**

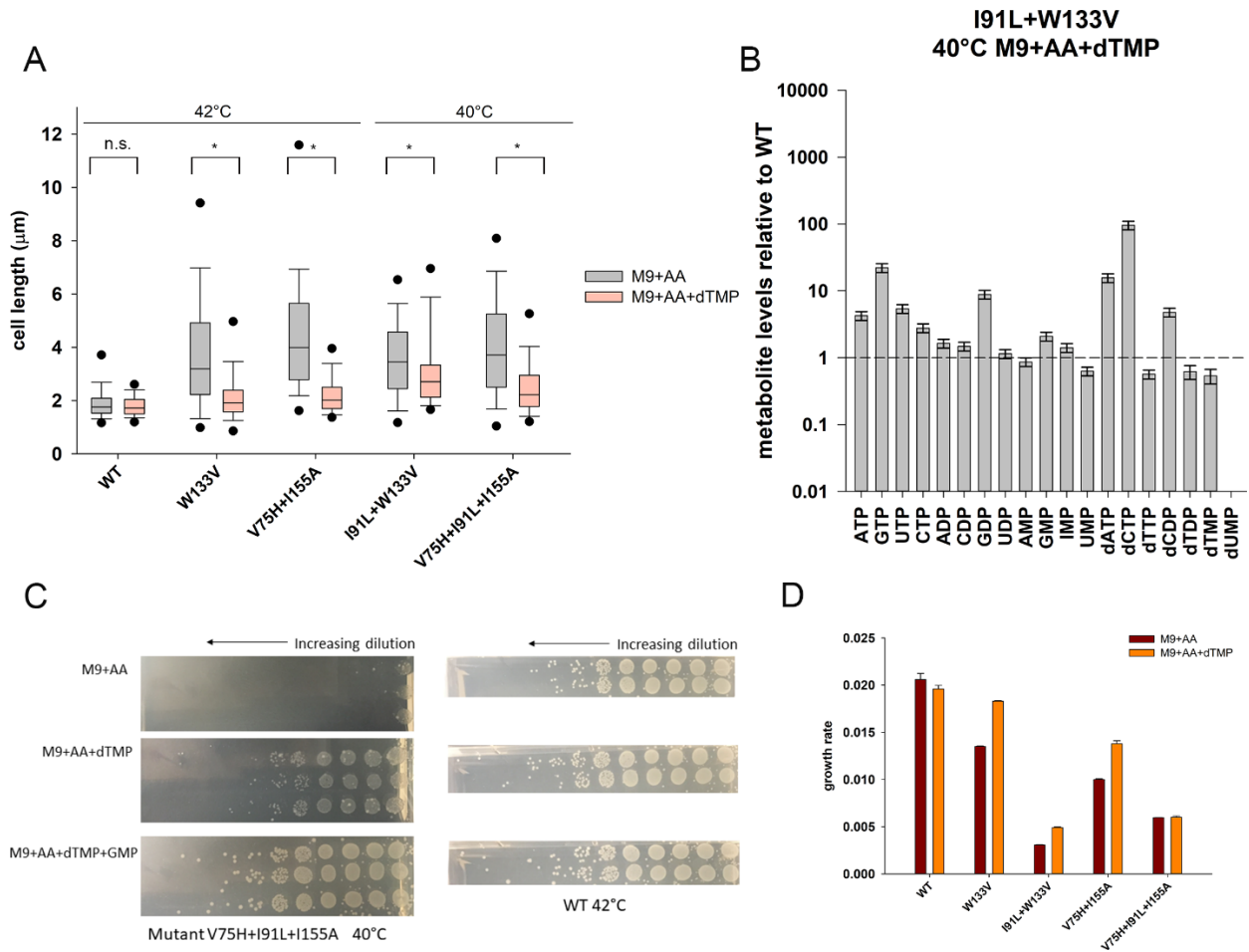

**Figure 6: Supplementation of dTMP alleviates dTDP/dTTP levels and rescues filamentation.** (A) Distribution of cell length of WT and mutant strains when grown in M9 medium supplemented with amino acids (gray) or with both amino acids and 1mM dTMP (pink). WT, W133V and V75H+I155A strains were grown at 42°C while I91L+W133V and V75H+I91L+I155A mutants were grown at 40°C. 1mM dTMP largely rescues filamentation of mutant strains (* indicates the median cell lengths were significantly different, Mann-Whitney test, p-value <0.001). (B) Abundances of selected nucleotides in I91L+W133V mutant when grown for 4 hours at 40°C in M9 medium supplemented with both amino acids and 1mM dTMP. Metabolite levels were normalized to those of WT grown under similar conditions. dTDP and dTTP levels recover and are now comparable to dTMP levels. Error bars represent SEM of 3 biological replicates. (C) Mutant V75H+ I91L+I155A grows very poorly (in terms of colony forming units, cfu) on a minimal media agar plate supplemented with amino acids (M9+AA) at 40°C. Supplementation of additional dTMP increases the cfu by several orders at the same temperature, while supplementation with both pyrimidine (dTMP) and purine (GMP) allows it to grow as good as WT. In comparison, WT was grown at 42°C under different supplementation conditions. In all cases, cultures were 7-fold serially diluted for the next spot. The three rows (two rows for WT) in each condition represent biological replicates. (D) Comparison of growth rates of WT and mutant DHFR strains at 42°C (40°C for I91L+W133V and V75H+I91L+I155A) in minimal medium that is supplemented with amino acids and/or 1mM dTMP. Except for W133V and to a lesser extent for V75H+I155A, the effect of dTMP on growth rates is only modest. Error bars represent SEM of 3 biological replicates.

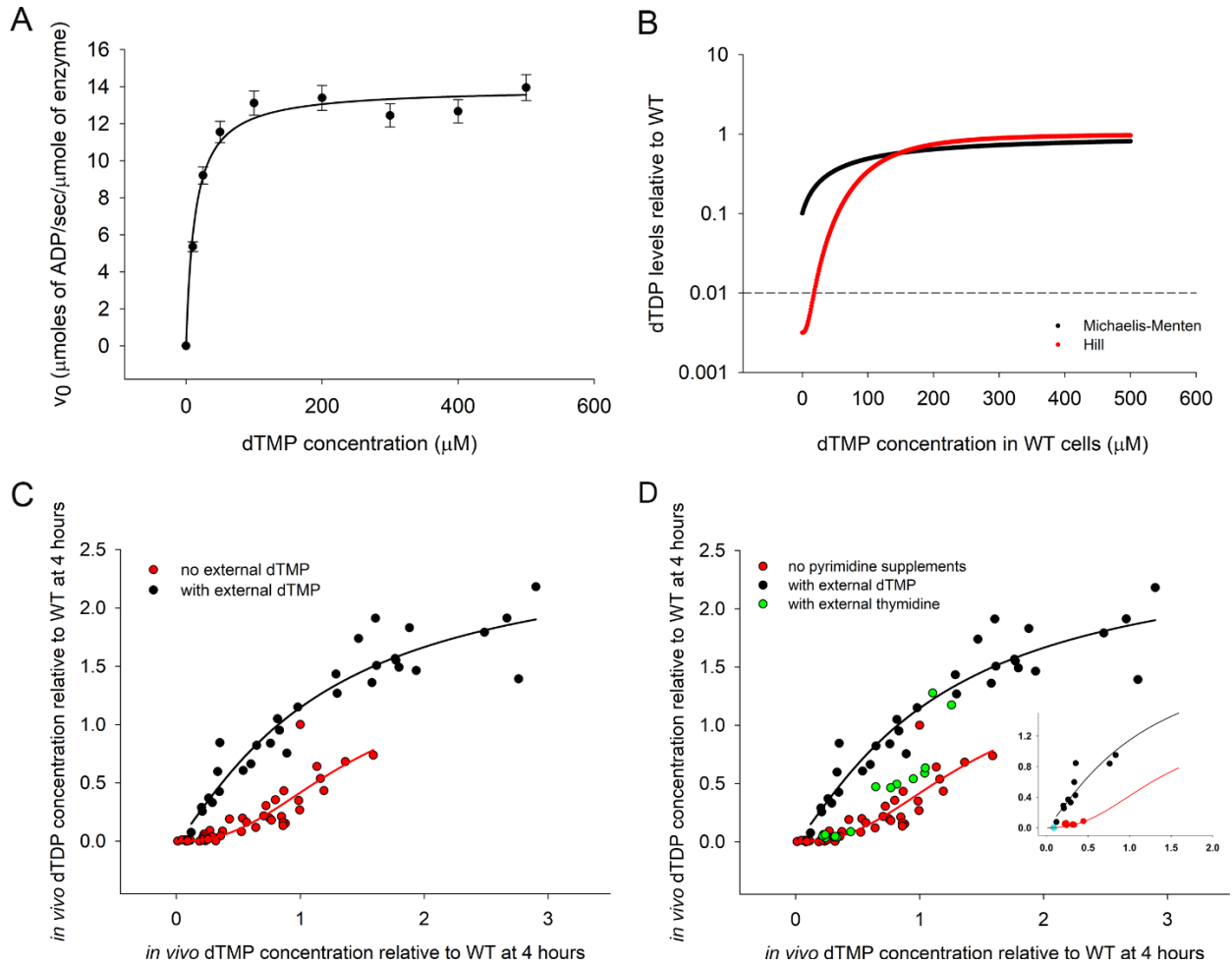

**Figure 7:** *In vivo* **enzyme kinetics of Tmk is highly cooperative and different from that *in vitro*.** (A) *in vitro* activity of purified Thymidylate Kinase (Tmk) as a function of dTMP concentration shows Michaelis-Menten (MM) like kinetics. ATP concentration is saturating at 1mM. The $K_M$ of dTMP is 13μM. (B) For a 10-fold drop in intracellular dTMP concentration in mutant relative to WT (as seen in I91L+W133V mutant), we calculate dTDP levels in mutant (relative to WT) as a function of various assumed intracellular concentrations of dTMP in WT (shown along x-axis), as absolute value of this is not known experimentally, assuming MM kinetics (black line) as shown in panel (A) or Hill kinetics with coefficient of 2.5 (red line) as shown in panel (C). The dotted line corresponds to the experimentally observed dTDP ratio of 0.01 for I91L+W133V mutant. This ratio is realized only for the Hill-like curve. (C) Apparent *in vivo* activity kinetics of Tmk enzyme using steady state dTMP and dTDP levels obtained from metabolomics. The plot includes data from WT, mutants W133V and I91L+W133V, as well as WT treated with 0.5µg/ml Trimethoprim, obtained at different time points during growth. Data points represent metabolite levels for all individual biological replicates without averaging. The black data points were acquired during growth in the presence of different concentrations of external dTMP (0.25, 0.5, 1, 2 and 5mM), while red points were from conditions with no external dTMP. Both black and red solid lines represent fit to Hill equation. (D) The green points were acquired during growth of WT, W133V and I91L+W133V mutants in the presence of different concentrations of thymidine, which follow the red curve. The inset plot shows the same graph with selected datapoints for I91L+W133V mutant. The cyan circle shows I91L+W133V mutant in the absence of any metabolite supplementation, while red and black points indicate metabolite levels following thymidine and dTMP supplementation respectively.

## Methods

**Strains and media.** The mutant DHFR strains chosen for this study (W133V, I91L+W133V, V75H+I155A and V75H+I91L+I155A) were a subset of strains generated in and described in Bershtein et al [20]. Briefly, using structural and sequence analyses, positions were chosen that were buried in the protein and located at least 4Å away from the active site, so that mutations introduced at these positions would have minimal effect on enzymatic activity. The mutations were intended to destabilize the protein as was confirmed by stability measurements of the purified proteins. The single mutants were in most cases mild to moderately destabilizing, and hence certain mutations were combined to increase the range of destabilization achieved. These mutations were eventually introduced into the chromosomal copy of the *folA* at its endogenous locus keeping its regulatory region intact, and the effect of the mutations on its growth and morphology were measured at 30, 37 and 42°C. The reason for choosing a wide range of temperature instead of following a single conventional temperature of 37°C was that *E. coli* is a gut bacterium and inhabit hosts whose core body temperatures span a large range (37-38°C for mammals, 40-45°C for birds [42,43]). Moreover, since the chosen mutants were temperature sensitive, the phenotypic manifestation of the mutations was the largest at the extremes of temperatures, in case of *E. coli* at 42ºC.

Wherever mentioned, M9 minimal medium *without amino acids* was only supplemented with 0.2% glucose and 1mM $MgSO_4$ while M9 media *with amino acids* was supplemented with 0.2% glucose, 1mM $MgSO_4$, 0.1% casamino acids, and 0.5 μg/ml thiamine. Casamino acids is a commercially available mixture of all amino acids except tryptophan, and cysteine is present in a very small amount. Wherever mentioned, 1mM GMP was used as a source of purine, while dTMP (thymidine monophosphate) was used at a concentration of 0.25-1mM. Thymidine was used at two different concentrations of 0.5mM and 1mM.

**Growth conditions.** All strains were grown overnight from a single colony at 30°C, and subsequently the culture was diluted to a final $OD_{600}$ of 0.01 in the specified medium and allowed to grow for 16-18 hours in Bioscreen C (Growth Curves, USA) at 30°C, 37°C or 42°C. Growth curves were fit to a 4-parameter Gompertz equation as described in [44] to derive growth parameters. Error bars were calculated as SEM of three biological replicates.

**Light microscopy.** Cells were grown overnight at 30°C from a single colony in the specified medium, diluted 1/100, and grown at various temperatures for 4 hours. For DIC images in Figures

1C, 1D, Figure S1, cells were pelleted, washed with PBS, and concentrated. DAPI staining (Molecular probes) was performed for 10 min at RT according to manufacturer instructions. 1 µl of a concentrated culture was then mounted on a slide and slightly pressed by a cover slip. DIC and DAPI images were obtained at room temperature by Nikon Ti Eclipse Microscope equipped with iXon EMCCD camera (Andor Technologies). For live phase contrast images and time-lapse experiments (Figure S8), cells were mounted on supplemented M9 + 1.5% low melting agarose (Calbiochem) pads. Pads were then flipped on #1.5 glass dish (Willco Wells), and the images were acquired at room temperature with Zeiss Cell Observer microscope. For DIC images in Figure S9, cells were placed on agar pads and images were acquired with Zeiss Cell Discoverer microscope.

**Analysis of cell lengths.** MicrobeTracker Suite [http://microbetracker.org/][45] was used to obtain distributions of cell length for phase contrast images and Zeiss Intellesis Module was used to analyze DIC images. On average, 500 cells were analyzed for each presented distribution. The cell lengths are represented in all figures as box-plots, where the boundaries of the box represent the $25^{th}$ and $75^{th}$ percentile, the line inside the box represents the median of the distribution, whiskers represent the $10^{th}$ and $90^{th}$ percentile, while the dots represent the $5^{th}$ and $95^{th}$ percentile.

**Statistical analysis.** In our experiments, cell lengths of *E. coli* were not normally distributed. Hence non-parametric Mann-Whitney test was used to determine if the median cell lengths of two samples were significantly different. In Figure 7C, fits to two different models, Michaelis-Menten and 3-parameter Hill were compared using extra sum-of-squares F-test using GraphPad Prism software v9.0.0.

**Metabolomics.** Cells were grown overnight at 30°C from a single colony in the specified medium, diluted 1/100, and re-grown. WT, WT+0.5µg/ml TMP and W133V mutant were grown at 42°C, while mutant I91L+W133V was grown at 40°C. For time course experiment, aliquots were removed after 2, 4, 6 and 8 hours, and metabolites were extracted as described in [15]. Briefly, the cells were washed 2 times with chilled 1×M9 salts, and metabolites were extracted using 300µl of 80:20 ratio of methanol:water that had been pre-chilled on dry ice. The cell suspension was immediately frozen in liquid nitrogen followed by a brief thawing (for 30 seconds) in a water bath maintained at 25°C and centrifugation at 4°C at maximum speed for 10 minutes. The supernatant was collected and stored on dry ice. This process of extraction of metabolite was repeated two more times. The final 900µl extract was spun down one more time and the supernatant was stored in -80°C till used for mass spectrometry. Metabolite levels were averaged over 2-3 biological replicates. In Figure 7C, data points represent metabolite levels for all biological replicates without averaging.

**Expression of SOS response genes by qPCR.** Cells were grown overnight at 30°C from a single colony in the specified medium, diluted 1/100, and grown at 37°C or 42°C for 4 hours. Based on $OD_{600}$ of the cultures, a volume equivalent to $5\times10^8$ cells were spun down (assuming 1 $OD_{600}$=$8\times10^8$ cells) and Protect Bacteria RNA Mini Kit (Qiagen) was used to extract total RNA as described in [16]. Following reverse transcription [16], expression of *recA*, *recN* and *sulA* genes were quantified using QuantiTect SYBR Green PCR kit (Qiagen) using the following primers:

recA_fwd   ACAAACAGAAAGCGTTGGCG
recA_rev   AGCGCGATATCCAGTGAAAG
recN_fwd   TTGGCACAACTGACCATCAG
recN_rev   GACCACCGAGACAAAGAC
sulA_fwd   GTACACTTCAGGCTATGCAC
sulA_rev   GCAACAGTAGAAGTTGCGTC

As it was difficult to find a reference gene that would be expressed to similar levels in WT vs mutant DHFR strains, we used total RNA to normalize the expression levels. Expression levels reported are average of 3 biological replicates. Error bars in Figure 4 represent 12% of the mean value.

**Tmk protein purification.** The tmk gene was cloned in pET28a plasmid between *NdeI* and *XhoI* sites with an N-terminal histag. BL21(DE3) cells transformed with the plasmid were grown in Luria Broth at 37°C till an OD of 0.6, induced using 1mM IPTG and grown for an additional 5 hours at 37°C. The protein was purified using Ni-NTA affinity columns (Qiagen) and subsequently purified by gel filtration using a HiLoad Superdex 75 pg column (GE). The protein was concentrated and stored in 10 mM potassium phosphate buffer (pH 7.2). The concentration of the proteins was measured by BCA assay (ThermoScientific) with BSA as standard.

**Tmk activity assay.** Tmk catalyzes the following reaction $dTMP + ATP \rightleftharpoons dTDP + ADP$, and the activity assay was carried out using the spectrophotometric assay as described in [25]. Briefly, the reaction mixture contained 5mM $MgCl_2$, 65mM KCl, 350uM phosphoenolpyruvate (PEP), and 300uM NADH. To obtain $K_M$ for dTMP, ATP concentration was fixed at 1mM, while dTMP concentration was varied from 10μM to 500μM. The reaction mix without enzymes was incubated at 25°C for 5 minutes, and the reaction was initiated by adding 100nM Tmk (final concentration) and 2 units of pyruvate/lactate dehydrogenase. The kinetic traces were recorded for every 5 seconds for a total time of 1 minute. The data corresponding to the first 20 seconds were fitted to a linear model to obtain initial rates. To obtain $K_I$ of dCTP for Tmk, ATP and dTMP concentrations were fixed at 100μM and 1mM respectively, while dCTP concentration was varied from 0.5mM to 7.5mM. Since conversion of dUMP to dTDP also produces ADP, the $K_I$ of dUMP could not be estimated by the spectrophotometric method. Instead, dTDP amounts produced in the reaction were determined by LC-MS. For the reaction, ATP and dTMP concentrations were fixed at 1mM and 100μM respectively, while dUMP concentration was varied from 0.25mM to 5mM. The reaction was quenched at 40 seconds using 80% MeOH. The resulting samples were subjected to LC-MS analysis to obtain dTDP levels. For Figure S7D, LC followed by mass spectrometry was used to directly measure dTDP levels.

**Data availability.** All metabolomics data for WT and mutants, as well as WT treated with Trimethoprim are included in Table S1.

## Supplemental Information

The manuscript contains 9 Supplementary figures (Figure S1-S9) and 1 Supplementary table.

**Table S1:** Metabolomics data

**Figure S1**

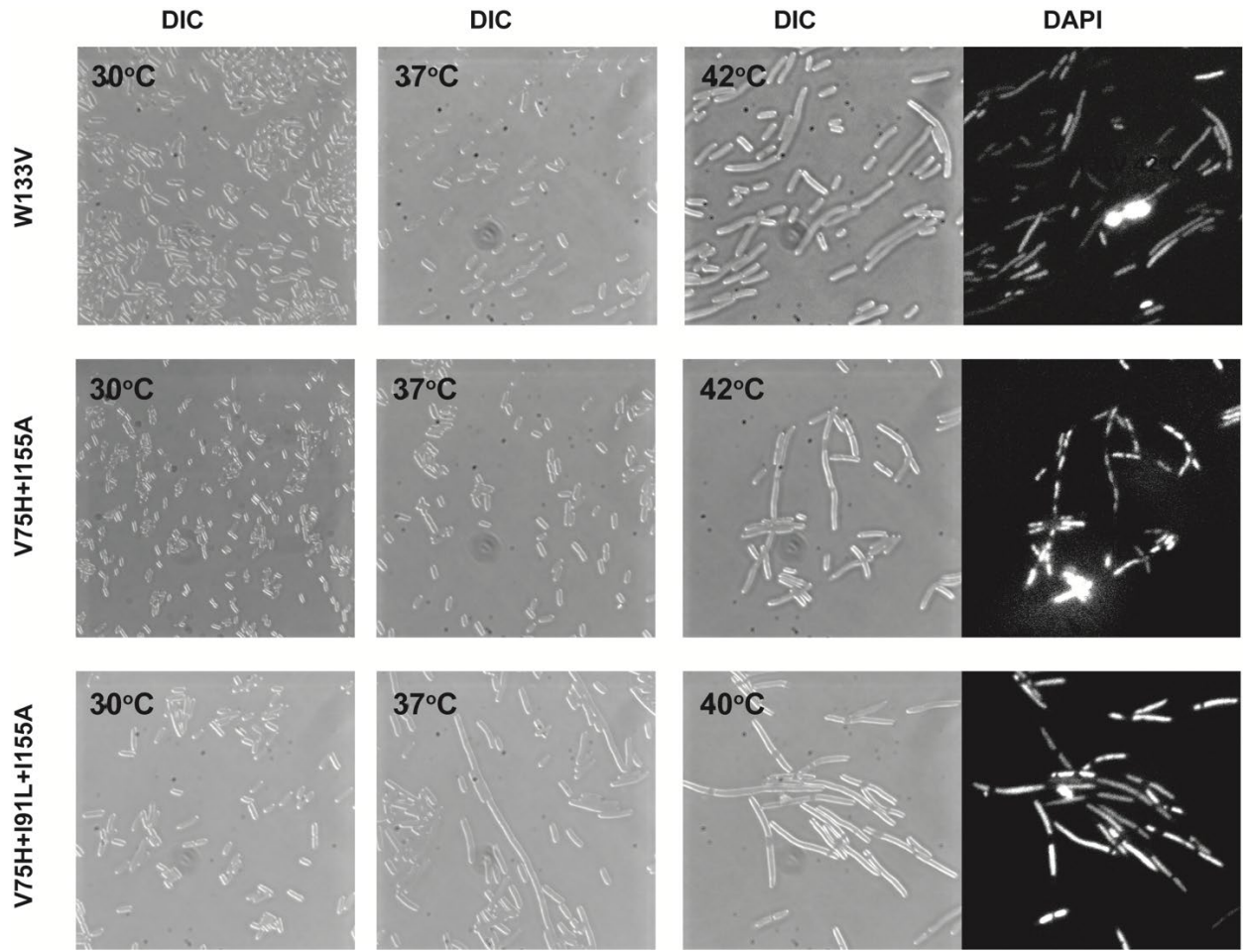

**Figure S1.** Destabilizing mutations in DHFR induce filamentous phenotype. Live cells DIC images with DAPI nucleoids staining of W133V, V75H+I155A, and V75H+I91L+I155A DHFR *E. coli* MG1655 strains. Prior to microscopy, cells were grown at 30°C, 37°C, and 42°C (V75H+I91L+I155A was grown at 40°C) in amino acid supplemented M9 medium for 4 hours (see *Materials and Methods*).

Figure S2

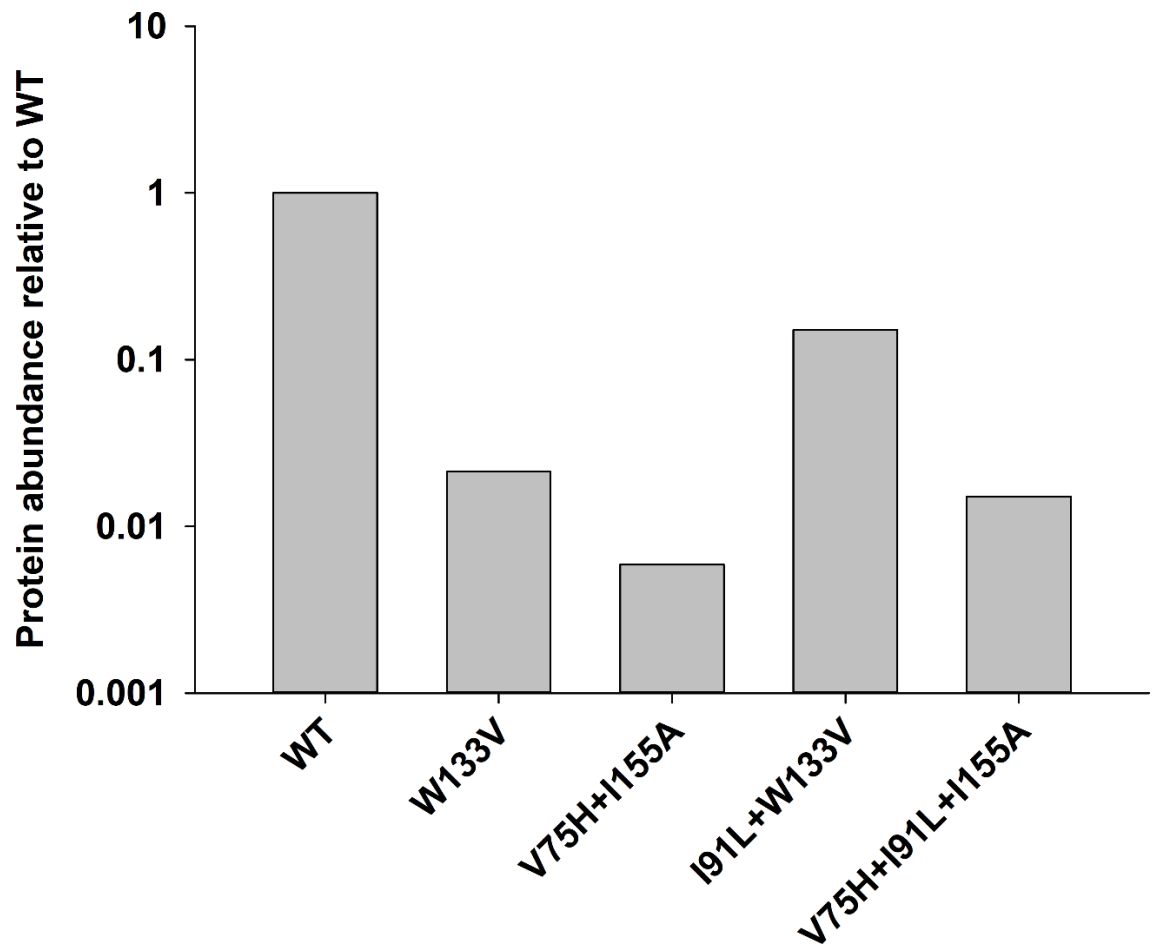

**Figure S2.** Intracellular abundance of WT and mutant DHFRs measured by Western blot. WT, W133V and V75H+I155A were grown for 4 hours at 42°C while I91L+W133V and V75H+I91L+I155A strains were grown for 4 hours at 40°C in amino acid supplemented M9 medium before being harvested. The data is also reported in [6].

**Figure S3**

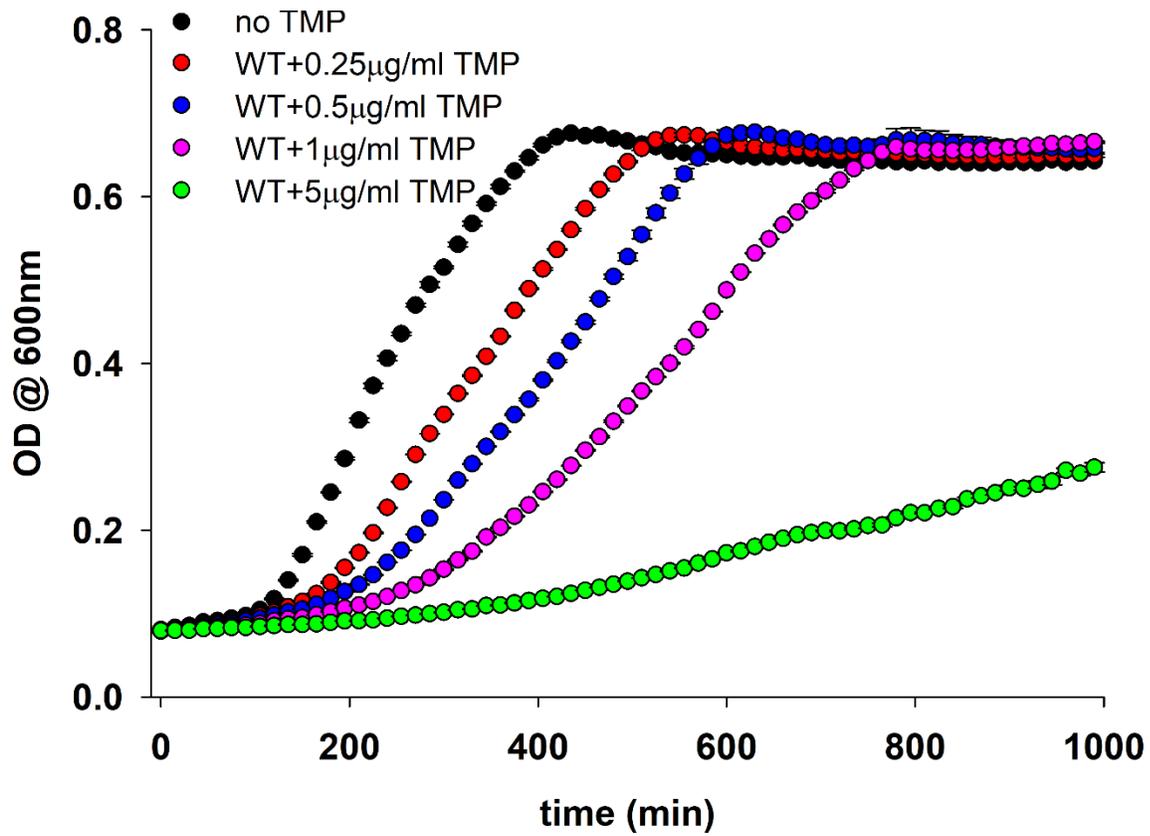

**Figure S3.** The effect of WT DHFR inhibition by trimethoprim (TMP) on growth. WT DHFR cells were grown at 42°C in amino acid M9 medium, and their growth was monitored by OD at 600nm. The data were fit to a 4-parameter Gompertz equation as described in [44] to derive growth parameters.

**Figure S4**

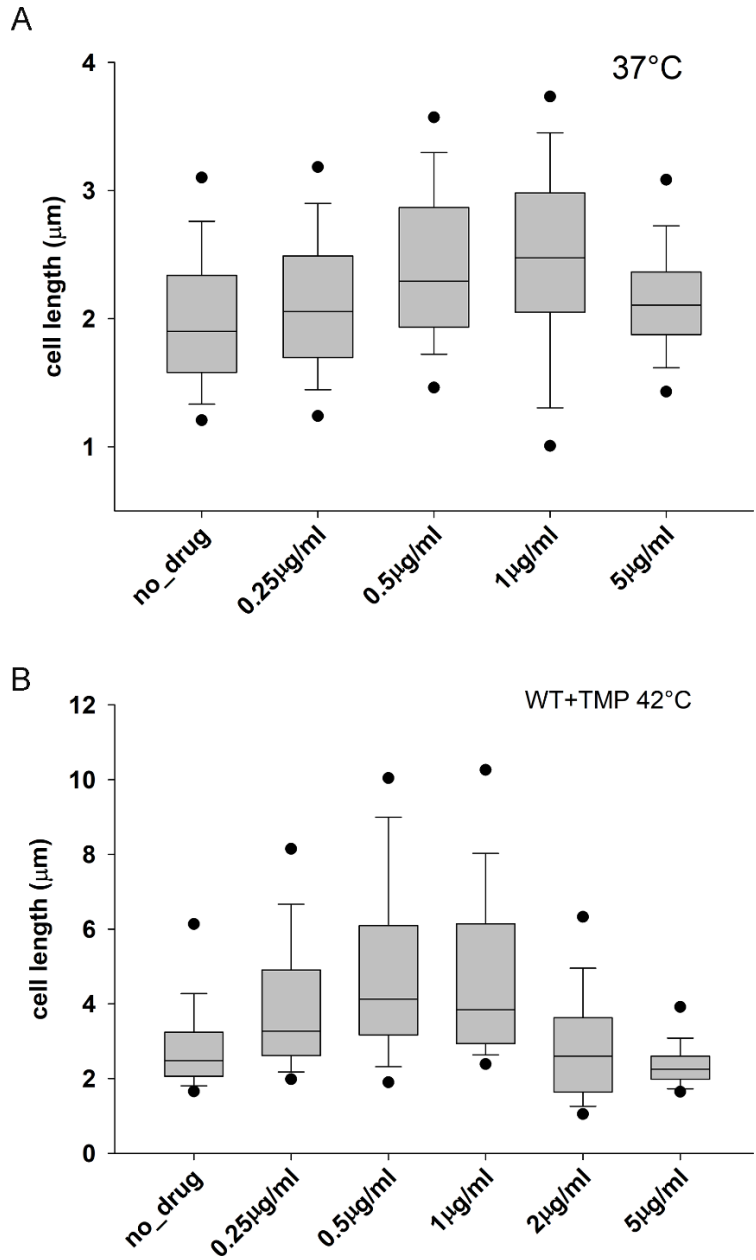

**Figure S4.** Distribution of cell length of WT *E. coli* as a function of TMP concentration when grown in amino acid supplemented M9 medium at (A) 37°C and (B) 42°C. Concentrations of TMP slightly below or near the MIC (1μg/ml) results in maximum filamentation, while the effect dies down at higher concentrations. Filamentation is much more pronounced at 42°C than at 37°C.

**Figure S5A**

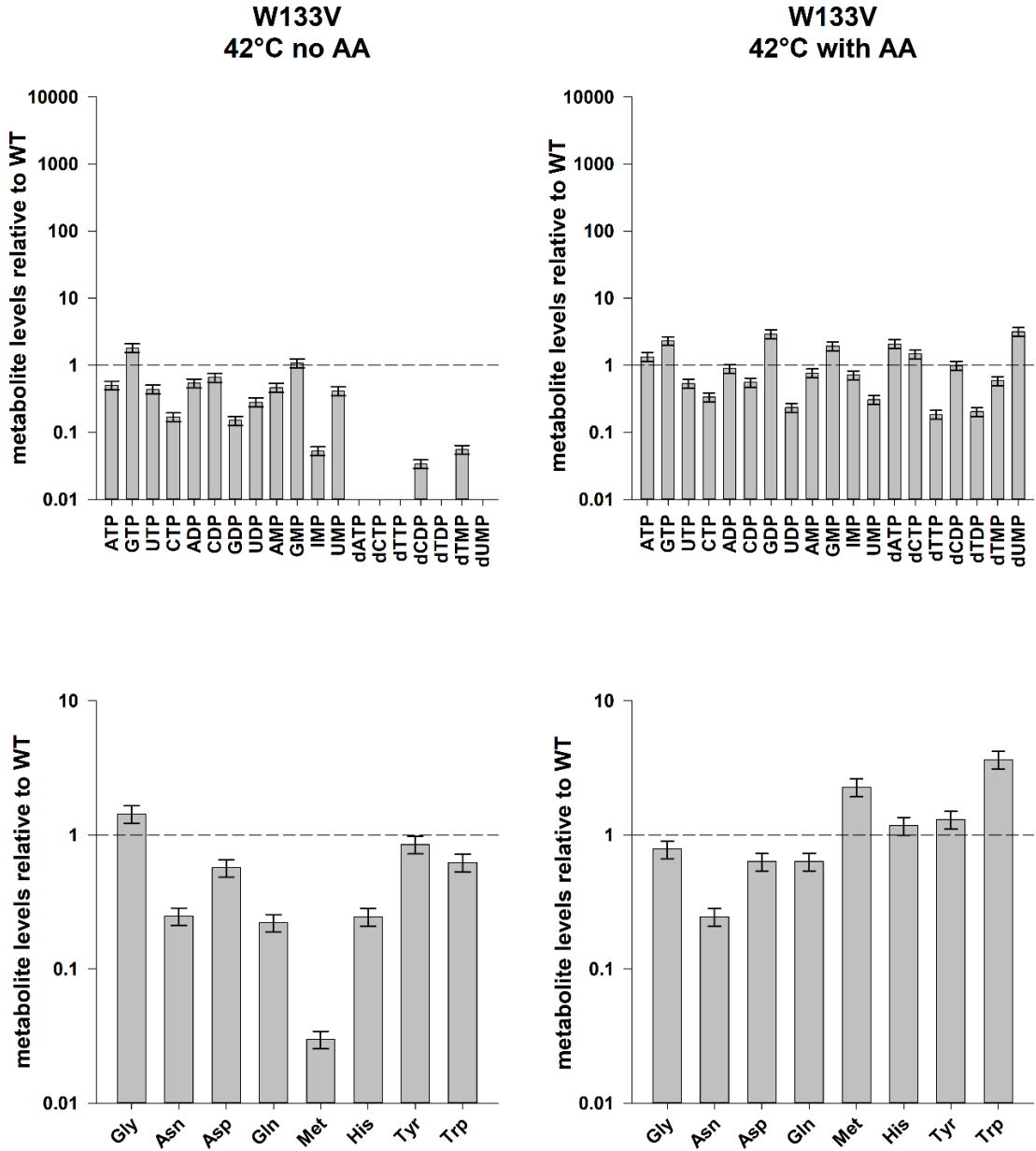

**Figure S5B**

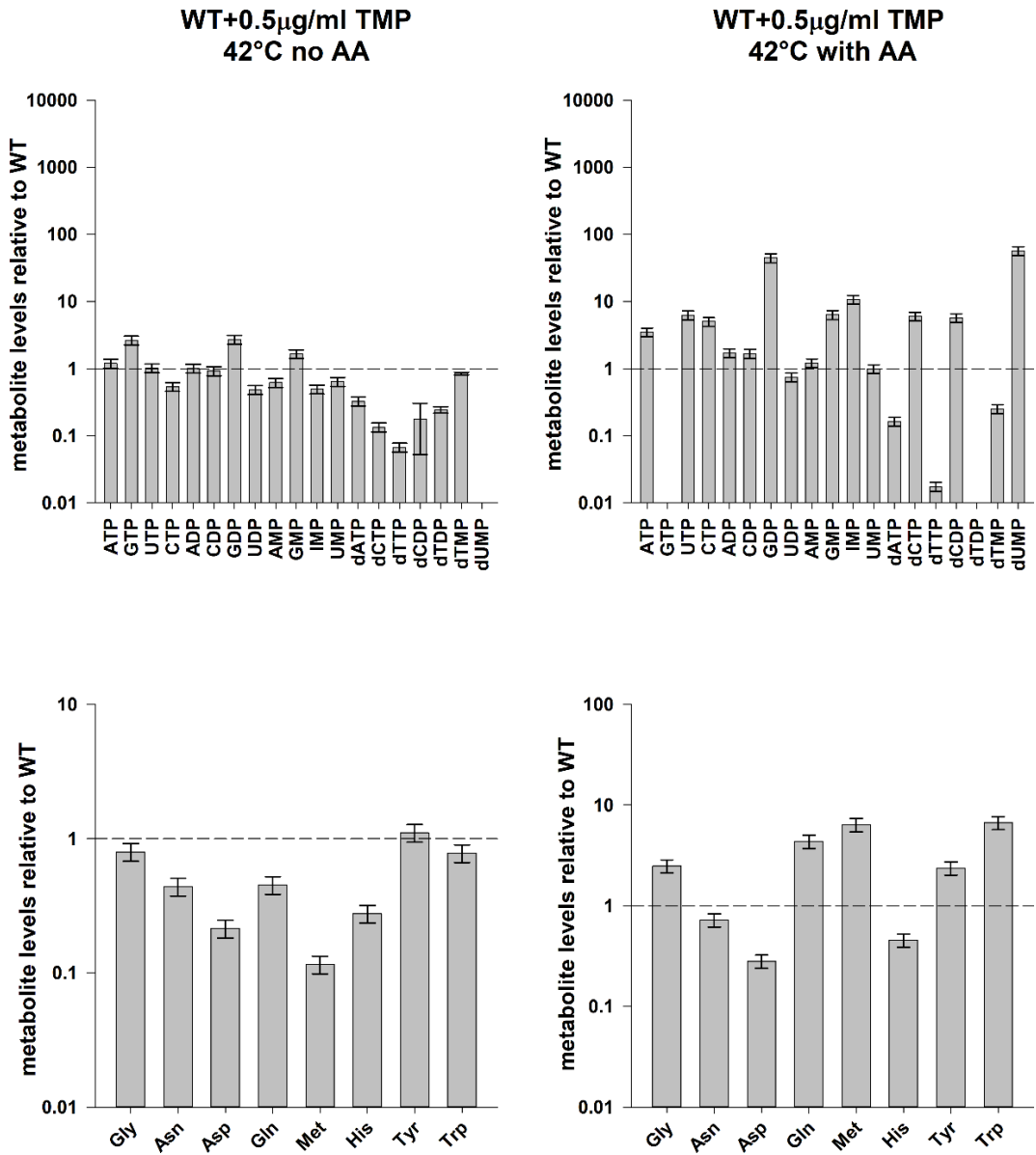

**Figure S5.** Metabolomics of (A) W133V DHFR strain and (B) WT treated with 0.5μg/ml Trimethoprim in minimal media at 42°C with or without added amino acids. The bars represent abundance of selected nucleotides and amino acids after 4 hours of growth in the indicated medium. Concentration of all metabolites were normalized to WT levels when grown under similar conditions. For both (A) and (B), methionine levels were extremely low in the absence of amino acids, which rise substantially when grown in the presence of amino acids. Though purines and pyrimidines also improve with amino acid supplementation, dTDP and dTTP levels remain poor.

**Figure S6**

A

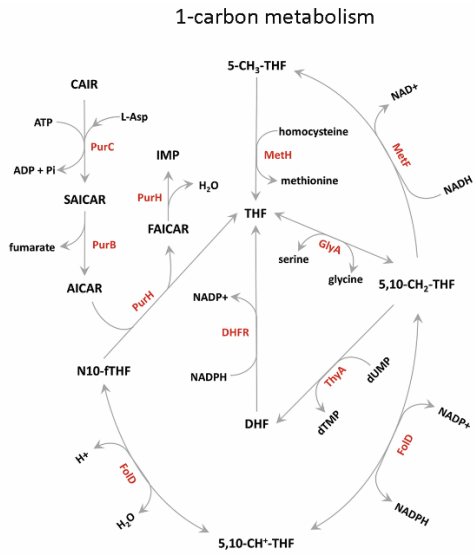

B

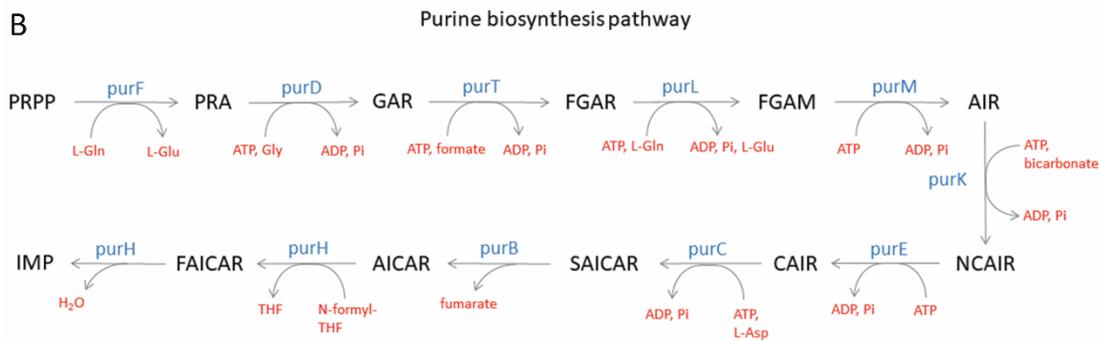

C

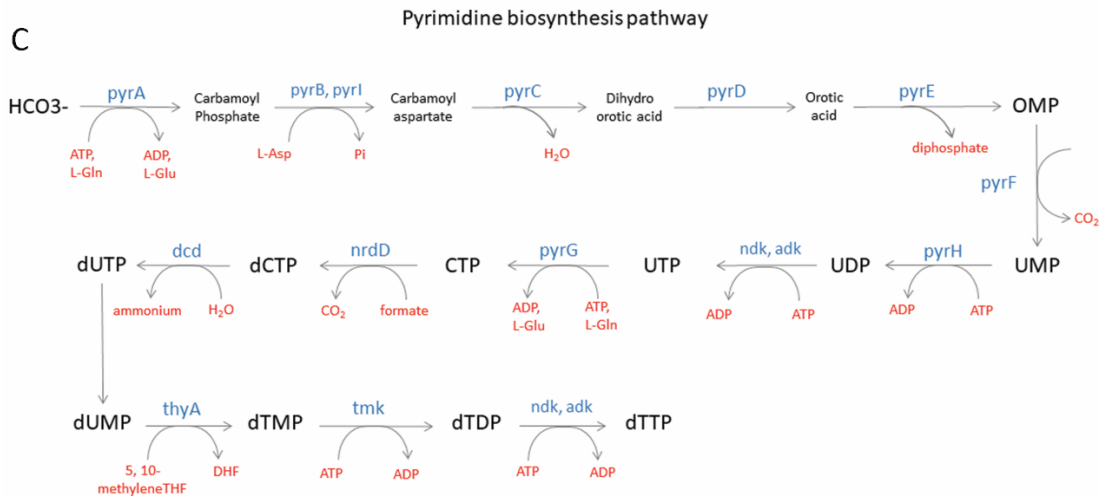

**Figure S6.** Schematic representation of (A) 1-carbon metabolism metabolism pathway (adapted from [15]) (B) *de novo* purine biosynthesis pathway and (C) *de novo* pyrimidine biosynthesis pathway.

**Figure S7**

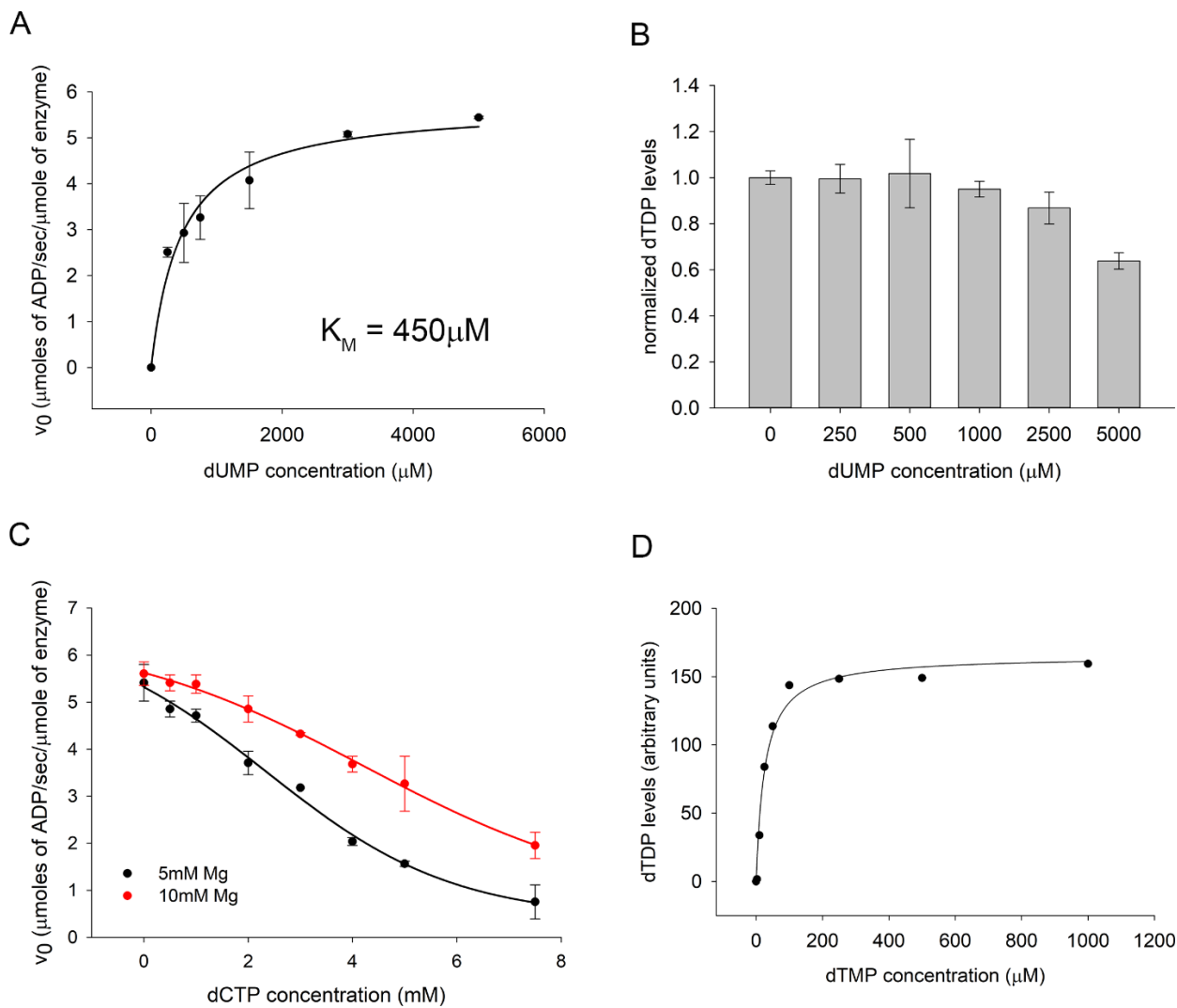

**Figure S7.** (A) Activity assay of purified Tmk enzyme using dUMP as substrate. ATP concentration is kept saturating at 1mM. The $K_M$ for dUMP is 450µM, compared to 13µM for dTMP. (B) Activity assay of Tmk was carried out in the presence of 100mM dTMP and 1mM ATP, and varying concentration of the inhibitor dUMP. dTDP levels were measured by HPLC followed by mass spectrometry. The data was fitted with a 4-parameter sigmoid curve to obtain an apparent $K_I$ of 3.9mM for dUMP. (C) Activity assay of Tmk was carried out in the presence of 100mM ATP and 1mM dTMP, and varying concentration of dCTP. ADP levels were measured using a NADH based coupled spectrophotometric assay. The red and black points indicate data acquired under different concentrations of $Mg^{2+}$. The data were fitted with a 4-parameter sigmoid

curve to obtain apparent $K_I$ of 2.3mM and 4.2mM at 5 and 10mM $Mg^{2+}$ concentrations respectively. (D) Activity assay of purified Tmk as a function of dTMP concentration in the presence of 5mM dUMP and 2.5mM dCTP as inhibitors. ATP concentration was kept saturating at 1mM. The dTDP levels were monitored using HPLC followed by mass-spectrometry. Even in the presence of inhibitors, the activity data here conforms to MM kinetics.

**Figure S8**

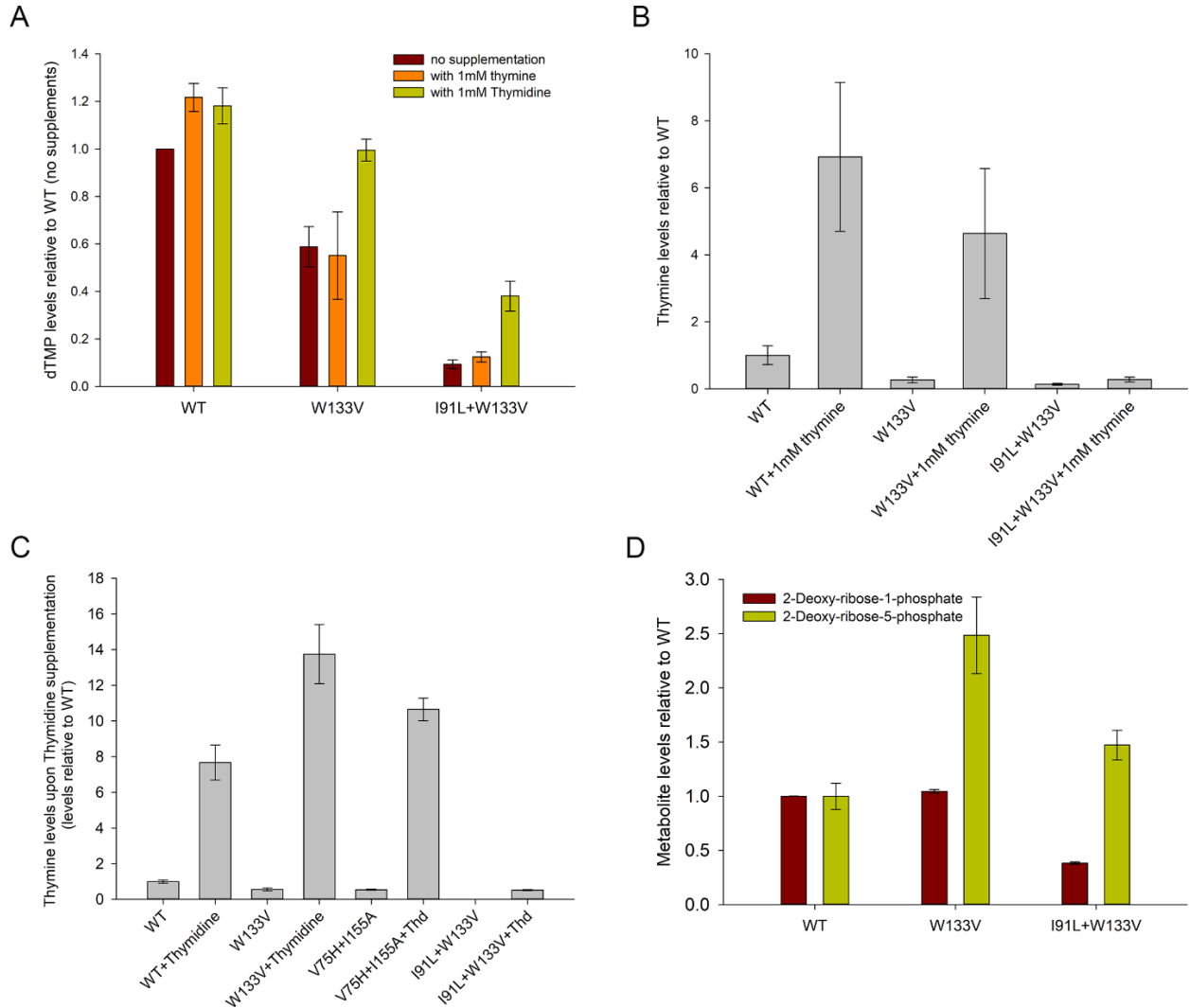

**Figure S8.** dTMP production through pyrimidine salvage pathway using thymidine and thymine supplementation. (A) Intracellular dTMP levels in WT and mutant strains upon addition of 1mM thymine or thymidine to the growth medium. Values are relative to those in WT strain (without any metabolite addition) after 4 hours of growth. Mutants show improvement in dTMP levels only upon thymidine addition. (B) Intracellular thymine levels in WT and mutant strains increase when grown in the presence of 1mM thymine in the medium, indicating that it is up taken by the cells. (C) Intracellular thymine levels in WT and mutant cells following growth with thymidine supplementation. Increase in thymine levels indicates substantial degradation of thymidine in the salvage pathway through DeoA enzyme. (D) Intracellular 2-deoxy-ribose-1-phosphate and 5-

phosphate levels in WT and mutant cells. Mutants accumulate substantially high levels of the 5-phosphate variant, indicating its channeling into energy metabolism.

**Figure S9**

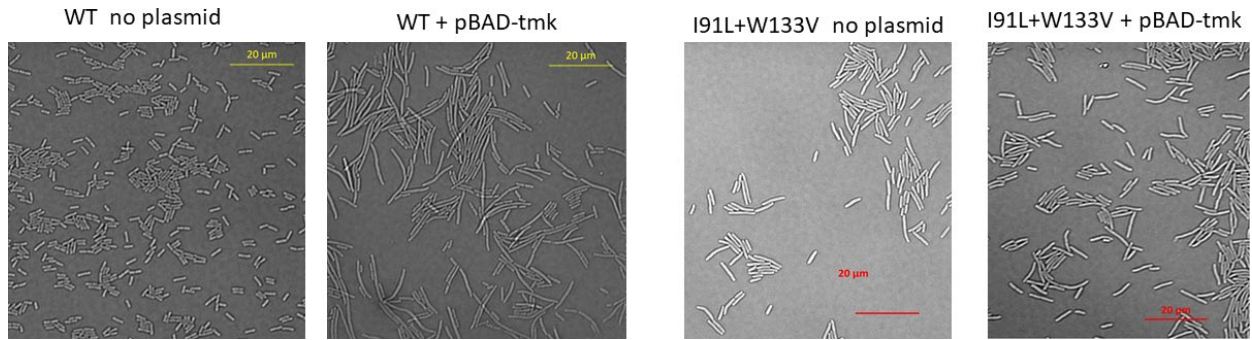

**Figure S9.** DIC images of untransformed WT and I91L+W133V mutant cells as well as those transformed with pBAD plasmid that expresses Thymidylate Kinase under control of arabinose promoter. Cells were grown at 42°C for 4 hours (40°C for mutant) in amino acid supplemented M9 medium in the presence of 0.2% of arabinose. While expression of Tmk does not rescue filamentation of mutant cells, it produces filamentation of WT cells.

# Supplementary Text

## I. Sequential enzymes in pathway with Michaelis-Menten kinetics

The following is an example of a pathway where several enzymes work sequentially:

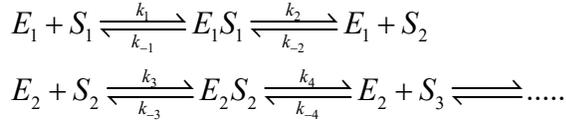

$$E_1 + S_1 \underset{k_{-1}}{\overset{k_1}{\rightleftharpoons}} E_1S_1 \underset{k_{-2}}{\overset{k_2}{\rightleftharpoons}} E_1 + S_2$$

$$E_2 + S_2 \underset{k_{-3}}{\overset{k_3}{\rightleftharpoons}} E_2S_2 \underset{k_{-4}}{\overset{k_4}{\rightleftharpoons}} E_2 + S_3 \rightleftharpoons \ldots$$

For the pathway at steady state, the concentrations of the reactants, products and intermediates do not change with time. Therefore,

$$\frac{d[E_1S_1]}{dt} = k_1[E_1][S_1] + k_{-2}[E_1][S_2] - (k_2 + k_{-1})[E_1S_1] = 0 \tag{4}$$

$$\frac{d[E_2S_2]}{dt} = k_3[E_2][S_2] + k_{-4}[E_2][S_3] - (k_4 + k_{-3})[E_2S_2] = 0 \tag{5}$$

The enzyme concentrations can be written as:

$$\begin{aligned}[E_1] &= [E_1]_0 - [E_1S_1] \\ [E_2] &= [E_2]_0 - [E_2S_2]\end{aligned} \tag{6}$$

$[S_1], [S_2]$ are the steady state concentrations of two sequential substrates (or products) in the pathway. Based on equations (1), (2) and (3), we deduce:

$$[E_1S_1] = \frac{(k_1[S_1] + k_{-2}[S_2])[E_1]_0}{k_{-1} + k_2 + k_1[S_1] + k_{-2}[S_2]} \tag{7}$$

$$[E_2S_2] = \frac{(k_3[S_2] + k_{-4}[S_3])[E_2]_0}{k_{-3} + k_4 + k_3[S_2] + k_{-4}[S_3]} \tag{8}$$

Again, at steady state, $\frac{d[E_1]}{dt} = 0 = -k_1[E_1][S_1] + k_{-1}[E_1S_1] - k_{-2}[E_1][S_2] + k_2[E_1S_1] \tag{9}$

Hence, $[E_1] = \frac{(k_2 + k_{-1})[E_1S_1]}{k_1[S_1] + k_{-2}[S_2]} \tag{10}$

Similarly, one can show that:

$$[E_2] = \frac{(k_4 + k_{-3})[E_2S_2]}{k_3[S_2] + k_{-4}[S_3]} \tag{11}$$

Since the pathway is at steady state, concentrations of reactants and products of every reaction remain unchanged with time, hence

$$\frac{d[S_2]}{dt} = k_2[E_1S_1] - k_{-2}[E_1][S_2] + k_{-3}[E_2S_2] - k_3[E_2][S_2] = 0 \quad (12)$$

Using the expressions of $[E_1], [E_2], [E_1S_1]$ and $[E_2S_2]$ from equations (4), (5), (7) and (8) into equation (9),

$$[E_1S_1]\left[k_2 - \frac{k_{-2}[S_2](k_2 + k_{-1})}{k_1[S_1] + k_{-2}[S_2]}\right] + [E_2S_2]\left[k_{-3} - \frac{k_3[S_2](k_4 + k_{-3})}{k_3[S_2] + k_{-4}[S_3]}\right] = 0$$

$$[E_1]_0 \frac{k_1 k_2 [S_1] - k_{-1} k_{-2} [S_2]}{k_{-1} + k_2 + k_1 [S_1] + k_{-2} [S_2]} = [E_2]_0 \frac{k_3 k_4 [S_2] - k_{-3} k_{-4} [S_3]}{k_{-3} + k_4 + k_3 [S_2] + k_{-4} [S_3]} \quad (13)$$

Using $K_{M1} = \frac{k_2 + k_{-1}}{k_1}, K_{M2'} = \frac{k_2 + k_{-1}}{k_{-2}}, K_{M2} = \frac{k_4 + k_{-3}}{k_3}, K_{M3} = \frac{k_4 + k_{-3}}{k_{-4}}$, where $K_{M1}$ is the Michaelis constant of $E_1$ for $S_1$, $K_{M2'}$ is that of $E_1$ for $S_2$, $K_{M2}$ is that of $E_2$ for $S_2$ and $K_{M3}$ is that of $E_2$ for $S_3$, equation (10) can be written as:

$$[E_1]_0 \frac{k_2[S_1]/K_{M1} - k_{-1}[S_2]/K_{M2'}}{1 + [S_1]/K_{M1} + [S_2]/K_{M2'}} = [E_2]_0 \frac{k_4[S_2]/K_{M2} - k_{-3}[S_3]/K_{M3}}{1 + [S_2]/K_{M2} + [S_3]/K_{M3}} \quad (14)$$

Assuming that $K_{M2'}, K_{M3} \gg K_{M1}, K_{M2}$ (in other words if $k_{-2}$ and $k_{-4}$ are very small) or the products have very low affinity back towards the enzyme, equation (11) reduces to the following:

$$[E_1]_0 \frac{k_2[S_1]}{K_{M1} + [S_1]} = [E_2]_0 \frac{k_4[S_2]}{K_{M2} + [S_2]} \quad (15)$$

Equation (12) can be re-arranged to get the following hyperbolic or Michaelis-Menten like dependence of $S_2$ on $S_1$:

$$[S_2] = \frac{k_2 K_{M2} [E_1]_0 [S_1]}{k_4 K_{M1} [E_2]_0 + (k_4 [E_2]_0 - k_2 [E_1]_0)[S_1]} = \frac{A[S_1]}{B + C[S_1]}, \quad (16)$$

Where $A = k_2 K_{M2}[E_1]_0, B = k_4 K_{M1}[E_2]_0, C = (k_4[E_2]_0 - k_2[E_1]_0)$

## II. Sequential enzymes in pathway with Hill-like kinetics

For a single enzyme, initial rate $v_0 = \dfrac{v_{max}[S]^m}{K_M + [S]^m}$, where '$m$' is the Hill coefficient

(1)

Now again consider the following scheme of sequential enzymes:

$$E_1 + S_1 \underset{k_{-1}}{\overset{k_1}{\rightleftharpoons}} E_1 S_1 \underset{k_{-2}}{\overset{k_2}{\rightleftharpoons}} E_1 + S_2$$

$$E_2 + S_2 \underset{k_{-3}}{\overset{k_3}{\rightleftharpoons}} E_2 S_2 \underset{k_{-4}}{\overset{k_4}{\rightleftharpoons}} E_2 + S_3 \rightleftharpoons .....$$

For reactants and products to be at steady state, earlier we derive Equation (12), which essentially equates the rate of production and consumption of $S_2$ through the two enzymes (with the assumption that $k_{-2}$ and $k_{-4}$ are very small). In such a situation if either $S_1$ or both $S_1$ and $S_2$ have limited diffusion, equation (12) can be written as the following based on equation (14):

$$[E_1]_0 \frac{k_2 [S_1]^m}{K^*_{M1} + [S_1]^m} = [E_2]_0 \frac{k_4 [S_2]^n}{K^*_{M2} + [S_2]^n} \quad (2)$$

Where m and n are the Hill coefficient analogs of the two consecutive enzymatic steps. Equation (15) can be rearranged as:

$$[S_2]^n = \frac{k_2 K^*_{M2} [E_1]_0 [S_1]^m}{k_4 K^*_{M1} [E_2]_0 + (k_4 [E_2]_0 - k_2 [E_1]_0)[S_1]^m} = \frac{A[S_1]^m}{B + C[S_1]^m} \quad (3)$$

Therefore, $[S_2] = \left[ \dfrac{A[S_1]^m}{B + C[S_1]^m} \right]^{1/n}$ (4)

A numerical solution of equation (17) shows that $S_2$ shows positive cooperativity as a function of $S_1$ only if $m > n$ (Figure below).

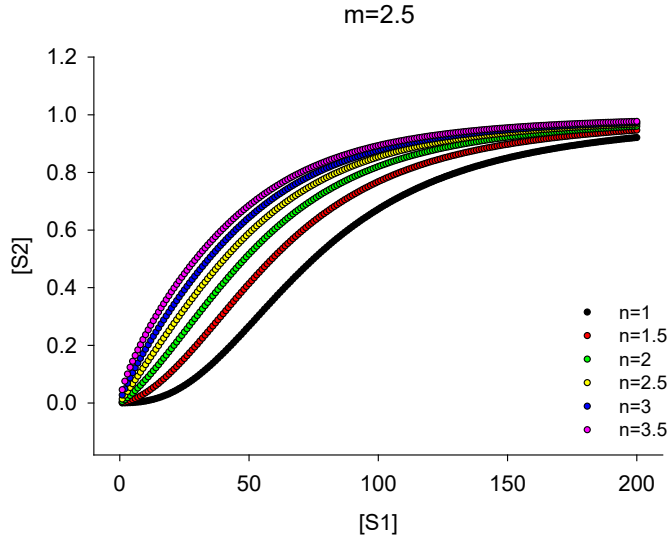

### III. Power law formalism for fractal kinetics

For a system where reactants and products diffuse freely, the rate constant of the reaction is time independent. However, under conditions of diffusion limitation (fractal kinetics), rate constant is no longer a constant, but varies with time in the following way:

$$k = k_0 t^{-h}$$

Where $h$ is related to the fractal dimension of the medium.

In the following section, we adopt the power law formalism as shown in [29] to convert the time dependent rate constant to a time independent one.

Considering the following simple reaction of two molecules of A forming a homodimer under conditions of diffusion limitation:

$$A + A \xrightarrow{k} product$$

$$rate = \frac{d[A]}{dt} = -k(t)[A]^2 = -k_0 t^{-h}[A]^2 \tag{1}$$

Integrating the above equation, we get [A] as a function of time

$$[A] = \frac{1-h}{k_0 t^{1-h}} \tag{2}$$

Rearranging this, we get

$$t = \left(\frac{1-h}{[A]k_0}\right)^{\frac{1}{1-h}} \tag{3}$$

In the next step, we replace $t$ in equation (1) with (3) to get:

$$Rate = -(1-h)^{\frac{h}{h-1}} k_0^{\frac{1}{1-h}} [A]^{2+\frac{h}{1-h}} = -k^*[A]^\alpha \tag{4}$$

Where $\alpha = 2 + \dfrac{h}{1-h}$ (Note that though this is a bimolecular reaction, the actual molecularity is >2 under fractal conditions)

**Application for enzyme kinetics**

Assume the following simple case:

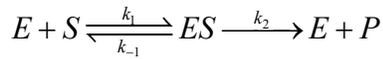

Consider that the substrate S is diffusion limited, hence $k_1$ (a bimolecular rate constant) will be time dependent.

$$\frac{d[ES]}{dt} = k_1(t)[E][S] - (k_{-1} + k_2)[ES] = k_1^*[E][S]^n - (k_{-1} + k_2)[ES]$$

Where $k_1^*$ is the apparent time independent rate constant, and $n$ is related to the fractal dimension of the medium.

At steady state, $\dfrac{d[ES]}{dt} = 0$

Hence, $[ES] = \dfrac{[E]_0 [S]^n}{\left(\dfrac{k_2 + k_{-1}}{k_1^*}\right) + [S]^n}$

Rate of the reaction $v = k_2[ES] = \dfrac{k_2[E]_0[S]^n}{\left(\dfrac{k_2 + k_{-1}}{k_1^*}\right) + [S]^n} = \dfrac{k_2[E]_0[S]^n}{K_M^* + [S]^n}$

Where $K_M^*$ is the apparent Michaelis constant, and $n$ is the Hill coefficient analog.